\documentclass[aps,prd,twocolumn,groupedaddress,showpacs]{revtex4-1}
\usepackage{amsmath,xfrac,graphicx,epstopdf,caption,subcaption,mathtools}
\usepackage{psfrag}

\begin{document}

\preprint{Draft 7: \today}

%Title of paper
\title{Sterile neutrino oscillations in core-collapse supernovae}

\author{MacKenzie L. Warren}
\email[]{mwarren3@nd.edu}
\author{Matthew Meixner}
\email[]{Matthew.Meixner@jhuapl.edu}
\author{Grant Mathews}
\email[]{gmathews@nd.edu}
%\homepage[]{Your web page}
%\thanks{}
%\altaffiliation{}
\affiliation{Department of Physics, Center for Astrophysics, University of Notre Dame, Notre Dame, Indiana 46556, USA}

\author{Jun Hidaka}
\email[]{jun.hidaka@meisei-u.ac.jp}
\affiliation{National Astronomical Observatory of Japan, 2-21-1 Osawa, Mitaka, Tokyo 181-8588, Japan}
\author{Toshitaka Kajino}
\email[]{kajino@nao.ac.jp}
\affiliation{National Astronomical Observatory of Japan, 2-21-1 Osawa, Mitaka, Tokyo 181-8588,  Japan\\
and\\
Department of Astronomy, Graduate School of Science, The
University of Tokyo, 7-3-1 Hongo, Bunkyo-ku, Tokyo, 113-0033, Japan}

\date{\today}

\begin{abstract}
We have made  core-collapse supernova simulations that allow  oscillations between electron neutrinos (or their antiparticles) with  right-handed sterile neutrinos.  We have considered a range of mixing angles and sterile neutrino masses including those consistent with sterile neutrinos as a dark matter candidate.  We examine whether such oscillations can impact the core bounce and shock reheating in supernovae.   We identify the optimum ranges of mixing angles and masses that can dramatically enhance the supernova explosion by efficiently transporting electron antineutrinos from the core to behind the shock, where they provide additional heating leading to much larger explosion kinetic energies.   We show that this effect can cause stars to explode that otherwise would have collapsed. We find that an interesting periodicity in the neutrino luminosity develops due to a cycle of depletion of the neutrino density by conversion to sterile neutrinos that shuts off the conversion, followed by a replenished neutrino density as neutrinos transport through the core.
\end{abstract}


\maketitle

\section{Introduction \label{sec:intro}}

Neutrino interactions play a significant role in the evolution of core-collapse supernovae due to the high matter densities and large neutrino fluxes achieved in this environment.  All three neutrino flavors are produced in roughly equal amounts during core-collapse supernovae (SNe) and the neutrino transport is sensitive to the details of neutrino mixing and interactions.  This makes supernovae a unique testing ground for neutrino physics.  

However, despite recent developments in computational methods and better understanding of the relevant micro-physics and hydrodynamics \cite{janka2012}, the detailed explosion mechanism for  core-collapse supernovae is still not known.  Nevertheless, whatever the explosion mechanism, it is clear the neutrinos play a very important role even if  new neutrino physics may ultimately be required.

One dilemma in the study of core-collapse supernovae is that most computational models in spherical symmetry do not successfully explode.  Also, although explosions occur more easily in  axisymmetric two-dimensional (2D) or fully three-dimensional (3D) models, the explosion energies tend to be too low~\cite{janka2012}.  

 The basic problem is that the initial outgoing shock loses energy due to the photodissociation of heavy nuclei in the outer core and becomes a standing accretion shock.  Recent efforts in supernova modeling have focused on revitalizing the explosion via  multidimensional hydrodynamic effects such as  the standing accretion shock instability \cite{blondin2003,blondin2006,scheck2008} or neutrino-heated convection~\cite{herant1994,burrows1995,janka1996,murphy2013}.  However, even in this case, the explosion energies are usually too low.  Clearly, the details of the neutrino transport and interactions are crucial to any currently proposed explosion scenario.  

In this context it is worth noting that even models with spherical symmetry can explode for progenitor stars of lowest mass \cite{kitaura2006,burrows2007} or by enhancing the flux of neutrinos emanating from the core via convection below the neutrinosphere
\cite{wilson1988,wilson1993,book,wilson2005}, or  from a second burst of neutrinos emanating from a QCD phase transition~\cite{fischer2011}.     In this paper we  explore yet another way in which even spherical models can explode  by  enhanced early neutrino luminosity.  In this case it occurs via the introduction of new  neutrino physics during the explosion. Specifically, we consider the possible resonant mixing \cite{hidaka2006,hidaka2007} between a sterile neutrino and an electron neutrino (or antineutrino).  

We find that, for a broad range  of sterile neutrino masses and mixing angles,  an efficient transport of antineutrino flux can occur by the resonant conversion of electron antineutrinos into  sterile neutrinos in the core followed by  the reverse process behind the shock.   We show that a substantial enhancement in the explosion can occur by this efficient transfer of energy out of the core.  Even models that otherwise do not explode can be made to explode by this process.  The heating occurs by the conversion of electron antineutrinos to sterile neutrinos.  The sterile neutrinos  then free stream out to just below  the neutrinosphere where they can convert back to normal electron antineutrinos.  These neutrinos  heat the protoneutron star surface just below the  neutrinosphere.  This heat  enhances the flux in both  electron antineutrinos and $\mu$, $\tau$ neutrinos that subsequently  contribute to heating behind  the shock.

Neutrinos are known to play a significant role \cite{bethe1985} in shock reheating.  The typical observed  supernova explosion kinetic energy is around $10^{51}$ ergs.  This is only $\sim$1\% of the energy released in  neutrinos during the first few seconds.  Reenergizing the shock clearly requires an understanding of the  detailed competition between neutrino heating and cooling \cite{book}.  Hence, any process that can increase the heating behind the shock can substantially enhance the explosion. 

Neutrino transport  also plays an important role in r process nucleosynthesis in neutrino driven winds \cite{woosley1994}.  However, current supernova models  \cite{fischer2010,hudepohl2010,wanajo2012} do not provide a sufficiently large neutron excess to produce a robust r process in this environment. 
New physics introduced by an oscillation to a sterile neutrino, however,  has the potential \cite{tamborra2012,wu2014} to help the  r process.  In this paper, however, we do not address the late times relevant to the r process.

The ``atmospheric" and ``solar" neutrino vacuum oscillation parameters have been well measured \cite{beringer2012}.  Depending upon mixing angles and matter densities achieved during the collapse, neutrinos in supernovae may also experience matter-enhanced Mikheyev-Smirnov-Wolfenstein (MSW) resonances \cite{wolfenstein1979}.  It has also been predicted \cite{kachelriess2005,serpico2012} that the mass hierarchy, whether ``normal" ($m_{1} < m_{2} < m_{3}$) or ``inverted" ($m_{3} < m_{1} < m_{2}$), will alter the emergent neutrino spectra if there is mixing and may even be detectable by the associated $\nu$-process nucleosynthesis in the C/He shell \cite{yoshida2008,mathews2012}.  In addition, neutrino self-interactions \cite{duan2010} may be important but are not as well understood due to their nonlinear nature.

In this paper, we investigate in particular the impact of a right-handed ``sterile" neutrino $\nu_{s}$ on the neutrino transport in core-collapse supernovae.  This proposed sterile electroweak singlet does not participate in the weak interactions and is therefore consistent with the Large Electron-Positron collaboration measurements \cite{LEP2006} of the width of the $Z^{0}$ gauge boson.  Indeed, the only directly detectable signature of a sterile neutrino might be via oscillations with the three active flavors.  

Nevertheless, bounds can be placed on the mass and mixing angle parameter space for a sterile neutrino by x-ray astronomy, cosmology  and supernovae \cite{abazajian2001a,boyarsky2006,chan2014}, i.e $1\text{ keV} < m_{s}< 18 \text{ keV}$ and $\sin^{2} 2 \theta_{2} < 1.93 \times 10^{-5} \left(\frac{m_{s}}{\text{keV}}\right)^{-5.04}$.  Recent measurements of x-ray emission from galaxy clusters are indicative (but not yet confirmed) of the possible presence of a 7.1~keV sterile neutrino dark matter candidate with $\sin^{2} 2 \theta \approx 7 \times 10^{-11}$  \cite{boyarsky2014,bulbul2014}.  As we shall see, however, this mixing angle is too small to produce  a significant effect in  supernova explosion dynamics via oscillations with active neutrinos.  Moreover, even if this result is correct the larger mixing angles of interest here are possible in other sterile neutrino flavors.  Alternatively, if the 7.1~keV sterile neutrino is only a component of the total dark matter, then a larger mixing angle could fit the observations~\cite{boyarsky2014,bulbul2014}.

There has already been much work on the effect of sterile neutrinos on the explosion mechanism and the r process in core-collapse supernovae \cite{caldwell2000,esmaili2014,fetter2003,hidaka2006,hidaka2007,mclaughlin1999,nunokawa1997,tamborra2012,wu2014}.  This was motivated by the anomalous results obtained at the Liquid Scintillator Neutrino Detector collaboration and other neutrino experiments that were better fit by (3+1) or (3+2) neutrino models.  However, most of these previous works utilized either analytic or parametrized models of the SN collapse and/or the neutrino driven wind environment.  In Ref.~\cite{wu2014} a study of the effect on explosion energy was made, but only for a very small $\sim 1$ eV sterile neutrino mass and a single mixing angle.  As we demonstrate below, this is outside the range of  sterile neutrino masses that would enhance the explosion (consistent with their result).   Hence, an investigation into the effects of a wide range of sterile neutrino masses and mixing angles within  a hydrodynamic  supernova model is warranted.  That is the main purpose of the work reported here.

Hidaka and Fuller \cite{hidaka2006,hidaka2007} proposed that a sterile neutrino with the mass and mixing angle of a warm dark matter candidate might significantly alter the dynamics of supernovae.  Using a one-zone collapse calculation, they found that a resonant conversion $\nu_{e} \rightarrow \nu_{s}$ could occur deep within the protoneutron star.  This initially decreases the energy of the shock by decreasing the electron fraction in the core.  However, a second resonant conversion $\nu_{s} \rightarrow \nu_{e}$ may occur just below the stalled shock.  Such a double resonance structure could enhance the neutrino-induced  reheating behind the shock by efficiently transporting the high energy electron neutrinos in the core out of the protoneutron star  to just below the stalled shock.  This mechanism, however, relies upon the detailed feedback of the neutrino oscillations onto the composition, energy transport, and hydrodynamics within the supernova environment.  It therefore requires a detailed numerical treatment, as we describe here.

In this paper, we explore    coherent active-to-sterile neutrino conversion.  We have incorporated  a sterile neutrino with a variety of masses and mixing angles into the University of Notre Dame/Lawrence Livermore National Laboratory (UND/LLNL) spherically symmetric relativistic supernova model~\cite{bowers1982,book}.   We then explore the impact of the sterile neutrino conversion on the shock heating.  We have considered sterile neutrinos with masses from  $10~\text{eV} \leq \Delta m_{s} \leq 10~\text{keV}$ and mixing angles in the interval $10^{-9} \leq \sin^{2} 2 \theta_{s} \leq 0.01$.  This range  encompasses  the dark matter constraints of Refs.~\cite{abazajian2001a,boyarsky2006}.  For a broad range  of sterile neutrino masses and mixing angles, we find that a substantial enhancement of the explosion can occur via the formation and subsequent reconversion of a sterile neutrino that  efficiently transfers neutrino heating from the core to the shock.

The organization of this paper is as follows.  In Sec.~\ref{sec:osc} we discuss the active-sterile MSW resonances. In Sec.~\ref{sec:model} we discuss the features of the UND/LLNL supernova model, and in Sec.~\ref{sec:method} we present the details of the  numerical treatment of the active-sterile neutrino oscillations.    The results of this work are presented in Sec~\ref{sec:results}, where we show that not only is an enhanced explosion possible, but an interesting cycle  develops in the neutrino luminosity due to the dynamics of the neutrino conversion process.  Our discussion and conclusions are provided in Sec~\ref{sec:conc}.

\section{Matter-Enhanced Sterile Neutrino Oscillations \label{sec:osc}}

With the inclusion of a sterile neutrino, vacuum oscillations require a full 4-neutrino flavor evolution.  For the present illustration, however, it is adequate to treat just the 2-neutrino mixing of $\nu_{e}-\nu_{s}$.  This is sufficient to explore the effects of a sterile neutrino on shock reheating  since electron neutrinos and antineutrinos play a dominant role in these phenomena.

The MSW mechanism \cite{mikheyev1985,wolfenstein1978} describes neutrino flavor mixing in matter, including supernovae.  As neutrinos propagate through matter, they experience an effective potential from charged and neutral current interactions due to forward scattering on baryonic and leptonic matter.  The forward scattering potential experienced by electron neutrinos in matter at some radius $r$ has the general form \cite{abazajian2001b}
\begin{eqnarray}
V(r) = &\sqrt{2} G_{F} \left((n_{e^-} - n_{e^+}) + 2 (n_{\nu_e} - n_{\bar{\nu}_e}) \right. \nonumber \\
&\left. (n_{\nu_\mu} - n_{\bar{\nu}_\mu}) + (n_{\nu_{\tau}} - n_{\bar{\nu}_\tau}) - n_{n}/2\right),
\label{eq:pot}
\end{eqnarray}
where $G_{F}$ is the Fermi coupling  constant and $n_{i}$ is the number density of species $i$.  This scattering potential derives from the asymmetries in matter and antimatter and depends upon the local matter composition.

In the supernova environment, it is safe to ignore the contribution from the forward scattering off of $\nu_{\mu,\tau}$ since the $\mu$ and $\tau$ neutrino and antineutrino flavors are created entirely by thermal pair production.  Therefore they arise in equal numbers.  Electron antineutrinos $\bar{\nu}_{e}$ are also created via thermal pair production processes; however, electron neutrinos are also generated by electron capture and, thus, a complete cancellation of $\nu_{e}$ and $\bar{\nu}_{e}$ does not occur.

The expression for $V(r)$ in Eq.~(\ref{eq:pot}) can be further simplified by using charge neutrality to relate the electron number density to the total proton density through $n_{p} = n_{e^-} - n_{e^+} = n_{B} Y_{e}$.  Also, the total neutron density can be written $n_{n} = n_{B} - n_{p}$.  After these simplifications, the forward scattering potential for electron neutrinos in supernovae reduces to
\begin{equation}
V(r) = \frac{3 \sqrt{2}}{2} G_{F}\, n_{B} \left( Y_{e} + \frac{4}{3} Y_{\nu_{e}} - \frac{1}{3}\right),
\label{eq:potential}
\end{equation}
where $n_{B}$ is the baryon number density.

The evolution of the forward scattering potential, and thus the neutrino flavor evolution, is determined entirely by the evolution of the baryon number density $n_{B}$, the electron fraction $Y_{e}$, and the electron neutrino fraction $Y_{\nu_e}$.  It is important to note that the oscillation of electron neutrinos into sterile neutrinos will in turn alter the local values of $Y_{e}$ and $Y_{\nu_e}$, allowing for feedback effects between the oscillations and the local hydrodynamic environment.

For a $\nu_{e} \rightarrow \nu_{s}$ conversion in medium, the vacuum oscillations are altered by matter and depend not only upon the local forward scattering potential $V(r)$, but also on the vacuum mixing parameters $\sin^{2} 2{\theta}$, the mass squared difference $\Delta m^{2}$, and the neutrino energy $E_{\nu}$.  An in-medium mixing angle \cite{wolfenstein1978} can be defined,
\begin{equation}
\sin^{2} 2 \theta_{M} (x) = \frac{\Delta^{2} \sin^{2} 2 \theta}{(\Delta \cos 2\theta - V(x))^{2} + \Delta^2 \sin^{2} 2\theta},
\end{equation}
where $\Delta = \Delta m^{2}/(2 E_{\nu})$.  Even if the vacuum mixing angle is small, it is possible to get maximal mixing in matter ($\theta_{M} = \pi/2$) if the condition $V(x) = \Delta \cos 2 \theta$ is satisfied.  This maximal mixing occurs at a MSW resonance.  For a given local environment, a MSW resonance will occur for a neutrino with energy
\begin{equation}
E_{res} = \frac{\Delta m^{2}}{2 V(r)} \cos{2 \theta}.
\label{eq:res}
\end{equation}
At  some radius  $r$ within the star, a neutrino with energy $E_{\nu}(r) = E_{res}$ will experience maximal mixing and undergo an oscillation to a sterile neutrino.  This resonance has a finite length scale (or time scale) along the neutrino's world line,
\begin{equation}
\Delta r_{res} = \left| \frac{d \ln V(r)}{dr}\right|^{-1} \tan 2 \theta_{s}.
\end{equation}
This corresponds to the distance over which the in-medium mixing falls to $\sin^{2} 2 \theta_{s} = 1/2$ and the forward scattering potential has changed by $\Delta V = \Delta m^2/(2 E_{res}) \sin 2\theta_{s}$.

There are two ways to induce an oscillation in medium: incoherent and coherent oscillations.  An incoherent conversion occurs when the neutrino mean free path $\lambda_{\nu}$ becomes short compared to the MSW resonance width $\Delta r_{res}$.  In this case scattering-induced incoherent conversion of the neutrino is enhanced by the presence of the resonance.  
Incoherent conversions will dominate when matter densities are large, such as in the center of the protoneutron star.  A coherent conversion will occur if the mean free path $\lambda_{\nu}$ is long compared to the resonance width $\Delta r_{res}$.  Here, we have considered only the effects of a coherent conversion of the neutrino flavor.  This is a good approximation because, for the sterile neutrino masses and mixing angles considered here, coherent  flavor evolution will dominate except for at the highest densities achieved during the core bounce  \cite{hidaka2006}.

In addition, in order to obtain a complete flavor conversion (i.e., all of the active neutrinos with the resonance energy $E_{res}$ oscillate to a sterile neutrino), one must force the conversion to be both coherent and adiabatic.  A coherent conversion can be ensured by requiring that the mean free path be much longer than the resonance width, $\Delta r_{res} \ll \lambda_{\nu}$.  The adiabaticity parameter $\gamma$ determines the efficiency of the conversion in a MSW resonance.  The adiabaticity parameter is the ratio of the resonance width $\Delta r_{res}$ to the oscillation length $\ell_{osc}^{res}$, i.e.,
\begin{equation}
\gamma = \frac{\Delta r_{res}}{\ell_{osc}^{res}},
\end{equation}
where
\begin{equation}
\ell_{osc}^{res} = \frac{4 \pi E_{res}}{\Delta m_{s}^{2} \sin 2 \theta_{s}}.
\end{equation}

A conversion will be adiabatic if there are many oscillation lengths within the resonance width, $\gamma \gg 1$, in which case the neutrino will completely switch flavors.  If both of these conditions are met, the flavor conversion is both coherent and adiabatic.  In that case, all of the electron neutrinos within the resonance energy width $\Delta E_{res} = \left(d E_{res}/dV\right) \Delta V \approx E_{res} \tan2 \theta$ around the resonance energy will oscillate to sterile neutrinos.

\section{Supernova Model \label{sec:model}}

Since our goal here is to find the change in SN explosion energy induced by resonant sterile neutrino oscillations, we begin with a model that can indeed explode.  For this we have a adopted the UND/LLNL  \cite{book} spherically symmetric general relativistic hydrodynamic core-collapse supernova model.  For ease of comparison with previous work, we start with the  20~M$_\odot$ progenitor model of~\cite{woosley1995}.  This model  explodes via enhanced early convection in the protoneutron star \cite{wilson2005}.   We note, however, that this enhanced early convection is not crucial to the active-sterile mixing effect, as we demonstrate below in Sec.~\ref{sec:results}.

The code utilizes a multigroup flux-limited diffusion (MGFLD) scheme (as described below) for the transport of three neutrino flavors ($\nu_{e}$, $\bar{\nu}_{e}$, and $\nu_{x}$).  Since $\mu$ and $\tau$ neutrinos and antineutrinos are created entirely from thermal pair production, it is acceptable for our purposes to treat them as a single $\nu_{x}$ flavor.  

For this work, we have used the equation of state (EOS) from references \cite{bowers1982,book}, which leads to a successful explosion.  We caution that the emergent neutrino spectra and the maximum mass of the resultant neutron star are sensitive to the EOS.  Even so, we will not consider other equations of state here. 

Our goal is simply to use this model as a fiducial starting place in which to judge the possible effects of mixing with a sterile neutrino.
Nevertheless, for completeness, we summarize here the basic physics and the way in which the effects of neutrino transport are implemented.  For details of the UND/LLNL supernova model, we refer the reader to Ref.~\cite{book}.

The metric used in the UND/LLNL general-relativistic supernova model is a variation of the May and White metric.  In Lagrangian coordinates we write,
\begin{align}
ds^2 = &- a^2 \left[1-\left(\frac{U}{\Gamma}\right)^{2} \right] dt^2 - \frac{2 a U}{\Gamma^2} \,dr \,dt + \frac{dr^2}{\Gamma^2} \nonumber \\
& + r^2 \left(d\theta^2+\sin^{2} \theta \, d\phi^2\right),
\label{eq:metric}
\end{align}
where $r$ is a coordinate distance and the metric coefficients are functions of mass $m$ and time $t$.  The coefficient $a$ is the inverse of the time component of the four velocity, $ a\equiv 1/U^{t}$, and is related to the gravitational redshift.  

The  metric  variable $\Gamma$ is defined as
\begin{equation}
\Gamma \equiv \left(1+U^{2} - \frac{2 M}{r}\right)^{1/2}.
\end{equation}
The quantity $M$ is the gravitational mass interior to radius $r$
 and $U^{2} = U^{r}U_{r}$ is the square of the radial component of the four velocity.

\subsection{Neutrino transport and flux-limited diffusion \label{sec:diff}}

Our neutrino spectra are defined such that $F_{\nu}(r,E)/E$ is the number density of neutrinos that lie in the energy group $\Delta E$ centered at energy $E$ at a zone located at a distance $r$.  This is related to the neutrino distribution function via
\begin{equation}
f_{\nu}(r,E) = 2 \pi^2 \left( \frac{\hbar c}{k}\right)^{3} \frac{F_{\nu}(r,E)}{E^{3}}.
\end{equation}

 The spatial evolution of the neutrino spectra $F_{i}$ for all six neutrino and antineutrino species are described by  a relativistic Boltzmann equation. 
 In the flux-limited diffusion approximation, the  Boltzmann equation is  simplified to a diffusion equation for the form
\begin{equation}
\frac{1}{a} \frac{\partial G_{i}}{\partial t} \approx \nabla \cdot \left(D_{i} \nabla G_{i}\right),
\label{eq:diff}
\end{equation}
 where $D_{i}$ is a flux-limited diffusion coefficient.  To achieve this form, one defines the angle-integrated  moments of the neutrino distribution function:
\begin{align}
G_{i} & = \int F_{i}\, d\Omega_{\nu} , \\
H_{i} & = \int F_{i} \cos \theta\, d \Omega_{\nu} ,\\
K_{i} & = \int F_{i} \cos^{2} \theta \,d\Omega_{\nu}.
\end{align}
Here,  $G_{i}$ is the angular-integrated energy distribution, $H_{i}$ is the flux, and $K_{i}$ is the pressure.  One can rewrite the neutrino transport equation in terms of these moments \cite{book},
\begin{align}
\frac{1}{a} \frac{\partial G_{i}}{\partial t} &+ q \frac{\partial G_{i}}{\partial q} \frac{1}{a} \frac{\partial}{\partial t} \left(\ln a\right) + \frac{1}{a r^{2}} \frac{1}{b} \frac{\partial}{\partial m} 
\left( a r^{2} H_{i}\right) \nonumber \\
& - \frac{U}{r} \left[ q \frac{\partial}{\partial q} \left(G_{i} - 3 K_{i}\right)\right] - \frac{1}{a} \frac{\partial}{\partial t} \left(\ln \rho \right) \left(G_{i} - q \frac{\partial K_{i}}{\partial q} \right) \nonumber\\
& = \int \Lambda_{i} \,d\Omega_{\nu},
\label{eq:bolt}
\end{align}
where $\int \Lambda_{i} d\Omega_{\nu}$ are the neutrino source and sink terms.  

Neutrino interactions include both charged- and neutral-current interactions for electron neutrinos and antineutrinos, with comparable neutral-current reactions for the $\mu$ and $\tau$ neutrinos.  Charged-current reactions are negligible for $\mu$ and $\tau$ neutrinos, however, due to the  small number of muons and tau particles present in the supernova environment.  One must also consider inelastic and coherent processes.  Most of these are described in more detail in Ref.~\cite{book}.

The angular degrees of freedom are integrated out on the right-hand side of Eq.~(\ref{eq:bolt}), so the neutrino   distribution functions $G_{i}$ become only functions of time, mass (or radial) coordinate, and neutrino total energy.    The source and sink terms $\Lambda_{i}$ are complicated functions computed from all of the relevant neutrino-matter and neutrino-neutrino interactions~\cite{book,mayle1985}.  Some of these interactions will be discussed below in Sec.~\ref{sec:int}.

 The concept of flux-limited diffusion is defined \cite{book} in terms of a dimensionless quantity
\begin{equation}
x \equiv \frac{\lambda \Gamma \left| \partial G_{i}/\partial r\right|}{G_{i}},
\end{equation}
where $\lambda$ is the neutrino mean free path.  When $x$ is small ($x \ll 1$), the angular distribution is isotropic.  In the limit of large $x$ ($x \gg 1$), the distribution is free streaming.  
In the diffusion limit, one assumes that the neutrino mean free path is small so that  the  radiative transfer equation reduces \cite{book} to 
\begin{equation}
H_{i} = -D_{i} \,\Gamma\, \frac{\partial G_{i}}{\partial r}.
\label{eq:flux}
\end{equation}

The method of flux-limited diffusion consists of finding a form for the diffusion coefficient $D_{i}$ such that Eq. (\ref{eq:diff}) remains valid from the diffusion limit (where the neutrino mean free path is small) to the case of neutrino free streaming over characteristic length scales of the simulation.  This can be achieved \cite{Mayle1988,book} by defining the flux-limited diffusion coefficient as follows:
\begin{equation}
D_{i} \approx \frac{\lambda_{i}}{3} \left(1+ h(x) \,x/3\right)^{-1}.
\label{eq:coeff}
\end{equation}
The quantity $h(x)$ is the flux limiter.  It is derived by constructing a Pad\'{e} series which fits an exact beam calculation of neutrino flow from high to low density regimes in steady state,
\begin{equation}
h(x) \approx \frac{4+x/2 + x^{2}/8}{1+x/2+x^{2}/8}.
\label{eq:limiter}
\end{equation}
Clearly, in the free streaming limit where $x\gg1$, $D\rightarrow G/(\lambda \Gamma) \left|\partial G/\partial r\right|$.  In the short mean free path limit $x \to 0$, $D\to \lambda/3$ as it should.  The neutrino flux calculated with Eqs. (\ref{eq:flux})-(\ref{eq:limiter}) generally agrees with exact  Boltzmann solutions to within a few percent \cite{wilson1993} for the form of $h(x)$ given in Eq.~(\ref{eq:limiter}).

\subsection{Multigroup neutrino energy transport \label{sec:int}}

 At high densities, the treatment of the neutrino spectrum is simplified because neutrinos are trapped and therefore well represented as Fermi-Dirac distribution functions.  However, for intermediate matter densities ($3 \times 10^{9}\, {\text{g}~\text{cm}^{-3} < \rho < 10^{12} \,{\text{g}~\text{cm}^{-3}}}$), the neutrinos can interact with matter but are not trapped.  In this transition region, 
 we treat the neutrino scattering as a multigroup spectral diffusion problem in the Fokker-Planck approximation \cite{wilson1975},
\begin{eqnarray}
\frac{dF_{i}}{dt} = &\epsilon \frac{\partial}{\partial \epsilon} \left\{ K_{i}\left[F_{i}(1-F_{i}/\alpha \epsilon_{\nu}^{3})\right.\right. \nonumber \\
&+kT\left.\left. \left(\frac{\partial F_{i}}{\partial \epsilon_{\nu}} - \frac{3 F_{i}}{\epsilon_{\nu}}\right)\right]\right\},
\label{eq:20}
\end{eqnarray}
where the diffusion coefficient $K_{i}$ is related to the relaxation time $\tau_{c}$ and the neutrino energy exchange per collision,
\begin{equation}
K_{i} = \left| \frac{\Delta \epsilon_{\nu}}{\epsilon_{\nu}}\right|_{coll} \tau_{c}^{-1}.
\end{equation}
The fractional energy loss per collision in this case is adjusted for the appropriate energy threshold \cite{bowers1982,book2,wilson1975,chang1970}.  The relaxation time is taken as
\begin{equation}
\tau_{c}^{-1} = \frac{n_{e} c \sigma_{e,i}}{1+ 2 n_{e} c \sigma_{e,i} \Delta t |\Delta \epsilon/\epsilon|_{coll}}
\end{equation}
where the electron-neutrino scattering cross section in Weinberg-Salam theory is given by $\sigma_{e,i} = c_{e,i}\sigma_{H}$, with
\begin{equation}
\sigma_{H} = \left\{ 
\begin{array}{l l}
7.66 \sigma_{0} \epsilon T, & T\leq \mu_{e} \\
0.98 \sigma_{0} \epsilon \mu_{e}, & T < \mu_{e} \text{ and } \epsilon > \mu_{e} \\
1.48 \sigma_{0} \epsilon \frac{T}{\mu_{e}} (1+11.6T/\epsilon)& \\
\hspace{0.5cm}\times (1+0.259\epsilon^{2}/T), & T<\mu_{e} \text{ and } \epsilon < \mu_{e}
\end{array}
\right.
\end{equation}
where $\sigma_{0} = 4 G^{2} m_{e}^{2}\hbar^{2}/\pi c^{2} = 1.7 \times 10^{-44} \,\text{cm}^{2}$ and $\epsilon$ is in MeV.  The numerical treatment of the Fokker-Planck equation is discussed in detail in Refs.~\cite{bowers1982,book}.  The finite difference form of Eq.~(\ref{eq:20}) reduces to the implicit solution of a diffusion equation.

Our model utilizes 
101 logarithmic energy groups \cite{bowers1982,book2} between 1 MeV and 200 MeV, i.e.,~energy group $j+1$ and $j+1/2$ grow as a power law.
\begin{equation}
E_{j+1} = \frac{E_2}{E_1}E_j,~~ E_{j+1/2} = \sqrt{\frac{E_2}{E_1}}E_j,
\label{eq:group}
\end{equation}
where $E_2/E_1 \approx 1.054$ is the ratio of energies of the first and second neutrino energy groups.  In this way the lowest energies are well sampled relative to the high energy tail of the neutrino distribution.

Recently it has been noted \cite{fischer2012,roberts2012} that the late time evolution of neutrino spectra and luminosities can be quite different due to a Boltzmann suppression factor from  the effective neutron-proton mass difference and Pauli blocking effects.  Nevertheless, since our goal here is to estimate the explosion energy at early times ($\sim500$~ms postbounce) the average neutrino energies are high and these subtle effects in the neutrino fluxes and energy spectra are not yet evident.  Hence,   the MGFLD approximations invoked here are adequate.  We expect that the results found here will therefore also occur in an exact Boltzmann solver.  This should, however,  ultimately be checked.

\section{Sterile Neutrino Transport \label{sec:method}}

To include a sterile neutrino in the model, resonance energies are easily calculated throughout the star using Eq.~(\ref{eq:res}) and the local matter composition at each time step.  The conditions for a coherent and adiabatic conversion can then be imposed.  If a resonance occurs, the number of neutrinos that oscillate between flavors must be determined from the neutrino spectra.  For a resonance at position $r$ and energy $E_{res}$, there will be $\Delta n_{\nu_{e} \rightarrow \nu_{s}} (r,E_{res})$ electron neutrinos that convert to sterile neutrinos and $\Delta n_{\nu_{s} \rightarrow \nu_{e}} (r,E_{res})$ sterile neutrinos that convert to electron neutrinos.  

The number of electron neutrinos that oscillate to sterile neutrinos will be given by the number of electron neutrinos that lie in the energy interval $E_{res} \pm \Delta E_{res} $, which can be found from the neutrino spectrum by integrating over this range,
\begin{equation}
\Delta n_{\nu_{e} \rightarrow \nu_{s}} = \int_{E_{res} - \Delta E_{res}}^{E_{res} + \Delta E_{res}} \frac{F_{\nu_{e}}(r,E)}{E} dE\;.
\end{equation}
Since $\Delta E_{res} \ll E_{res}$ for small mixing angles and $\Delta E_{res}$ is much smaller than the energy group widths, this integral can be approximated as 
\begin{equation}
\Delta n_{\nu_{e} \rightarrow \nu_{s}} \approx F_{\nu_{e}} (r,E_{res}) \frac{2 \Delta E_{res}}{E_{res}}.
\end{equation}
A similar argument can be made for the number of sterile neutrinos that oscillate to electron neutrinos $\Delta n_{\nu_{s} \rightarrow \nu_{e}}$.  Thus, the number of electron neutrinos left after encountering a resonance is 
\begin{align}
n'_{\nu_{e}} (r,E_{res}) = \;&n_{\nu_{e}} (r,E_{res}) - \Delta n_{\nu_{e}\rightarrow \nu_{s}} (r,E_{res})\nonumber \\
& + \Delta n_{\nu_{s}\rightarrow \nu_{e}} (r,E_{res})
\end{align}
and the net number of sterile neutrinos produced is
\begin{align}
n'_{\nu_{s}}(r,E_{res}) = \; &n_{\nu_{s}} (r,E_{res}) - \Delta n_{\nu_{s} \rightarrow \nu_{e}} (r,E_{res})\nonumber \\
& + \Delta n_{\nu_{e}\rightarrow \nu_{s}} (r,E_{res}).
\end{align}

The spectral intensity in sterile neutrinos can then be found from the relation between number density and spectral density:
\begin{equation}
n'_{\nu_s}  = \int_{E_j}^{E_{j+1}} \frac{F'_{\nu_s}}{E} dE \approx F'_{\nu_s}\frac{E_{j+1} - E_j}{E_{j+1/2}},
\label{specint1}
\end{equation}
where $j$ is the energy group number.  Again, this  approximation to the integral is valid since the energy group  widths are much smaller than the neutrino energies.  Inverting Eq.~(\ref{specint1}) we then have 
% we can construct  the sterile neutrino spectra $F'_{\nu_s}$ from the sterile-neutrino number densities using the approximation
\begin{equation}
F'_{\nu_s}(k,j) \approx n'_{\nu_s}  \frac{E_{j+1/2}}{E_{j+1} - E_j} = \frac{n'_{\nu_s}(k,j)}{\sqrt{{E_{2}}/{E_{1}}} - \sqrt{E_{1}/{E_{2}}}},
\end{equation}
where  $k$ is the radial zone number, and  the second equality follows from the way in which we set up the energy groups [cf. Eq.~(\ref{eq:group})].

Once sterile neutrinos have been generated through oscillations, they must also be allowed to propagate through the star.  It is unnecessary to implement a flux-limited diffusion scheme for the sterile neutrinos because one can assume that they are born in the ``free streaming" limit and that $E_{\nu} \gg m_{\nu}$, so the neutrinos propagate outward at near the speed of light.  In this case, the transport scheme involves no energy diffusion and the flux only diffuses radially as
\begin{equation}
F_{\nu_{s}}(r')  = \left(\frac{r}{r'}\right)^2 F_{\nu_{s}}(r).
\end{equation}

Light sterile neutrinos will stream out of the star until they encounter a second resonance and oscillate back to electron neutrinos.  In order to properly account for the possibility of a second resonance at a distance less than $c \Delta t$, it is necessary to compute the transport of the sterile neutrinos during the same time step as their creation.  This is a reasonable approximation since the resonance energies do not change drastically with each time step.  If no secondary resonance is encountered, the sterile neutrinos are allowed to free stream for  a distance $c\Delta t$ before the next time step.

\section{Results \label{sec:results}}

Simulations were run for a range of sterile neutrino masses ($10~\text{eV} \leq \Delta m_{s} \leq 10~\text{keV}$) and mixing angles ($10^{-9}\leq \sin^{2} 2\theta_{s}\leq 0.02$).  This parameter space encompasses masses and mixing angles consistent with the recent anomalous reactor results as well as dark matter candidates.  All of the simulations were done using the solar metallicity  $20\, \text{M}_{\odot}$ progenitor model of \cite{woosley1994} evolved with 249 radial logarithmic zones and 101 neutrino energy groups [cf.~Eq.~(\ref{eq:group})].

\begin{figure}[t!]
   \centering
   \includegraphics[width = 0.5\textwidth]{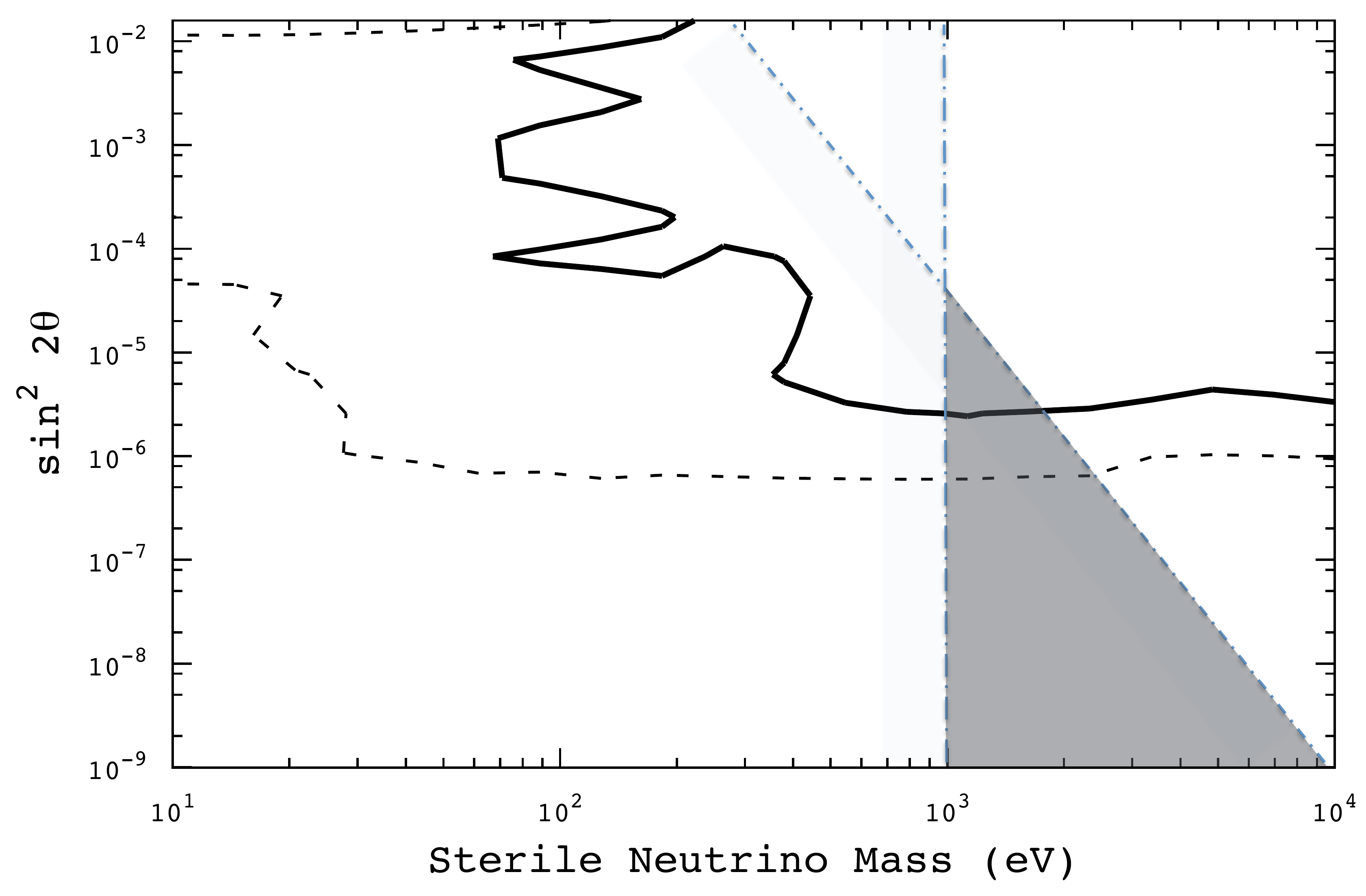} % requires the graphicx package
   \caption{Contours showing the enhancement in the kinetic energy at 550~ms postbounce, relative to an explosion without a sterile neutrino present.  The dashed line encloses the region of a factor of $1.5\,\times$ enhancement to the kinetic energy and the solid line encloses the region of a factor of $10\,\times$ enhancement.  The shaded region indicates the parameter space consistent with dark matter candidates \cite{abazajian2001b,hidaka2007}.}
   \label{fig:fig1}
\end{figure}

Figure \ref{fig:fig1} shows contours of  the total kinetic energy relative to an explosion without a sterile neutrino, at 0.55~s postbounce.  There is a substantial region of the parameter space that enhances the shock energies by at least a few percent and up to factors of 10.  This region of the parameter space is  consistent with  most cosmological dark matter constraints as indicated by the shaded lines on the figure.

Further insight is to be gained by analyzing in detail the evolution of  an explosion model with $\Delta m_{s}$ and $\sin^{2} 2 \theta_{s} $ selected from  the region of enhanced explosion energy in Fig~\ref{fig:fig1}.

\subsection{Example of an enhanced explosion: $\Delta m_{s} = 5.012$ keV and  $\sin^{2} 2 \theta_{s} =  1.12\times 10^{-5}$}

\subsubsection{Analysis of shock enhancement  \label{sec:shock}}

\begin{figure}[t!]
   \centering
   \includegraphics[width = 0.5 \textwidth]{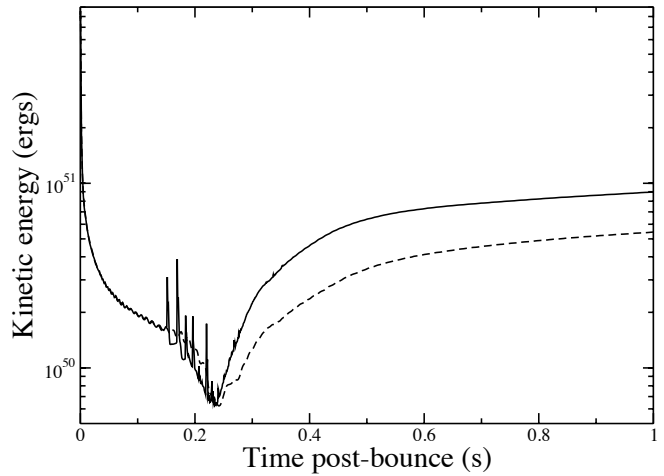} % requires the graphicx package
   \caption{Kinetic energy versus time postbounce.  The solid line is for a model including sterile neutrinos with $\Delta m_{s} = 5.012$ keV and  $\sin^{2} 2 \theta_{s} =  1.12\times 10^{-5}$ and the dashed line is for a model without a sterile neutrino.  The kinetic energy is enhanced with the oscillation into a  sterile neutrino.}
   \label{fig:fig2}
\end{figure}

As an example of a modestly  enhanced explosion, Fig.~\ref{fig:fig2} shows kinetic energy versus time for a sterile neutrino near the center-right side  of Fig.~\ref{fig:fig1} with $\Delta m_{s} = 5.012$ keV and  $\sin^{2} 2 \theta_{s} =  1.12\times 10^{-5}$.  This figure shows an initial decrease in total kinetic energy as material falls onto the forming protoneutron star.  However, the total kinetic energy rapidly increases after about 200 ms due to the delayed heating of material by neutrinos.  This causes the formation of a neutrino-heated bubble and the launch of material from the star.  

The presence of the sterile neutrino leads to an enhancement of the kinetic energy by about a factor of 1.5 as   shown in Fig.~\ref{fig:fig1} and  \ref{fig:fig2}.  Starting at about 200~ms postbounce, the kinetic energy with a sterile neutrino present exceeds that without a sterile neutrino.  Specifically, the kinetic energy is enhanced by a factor of $1.79$ at 0.55~s postbounce and remains  enhanced by a factor of $1.27$ at 9.55~s postbounce.   Note, that although there is a strong enhancement in the kinetic energy with this choice of mass and mixing angle, the kinetic energy of the explosion does not exceed observed supernova energies of $\sim$ a few $\times \,10^{51}$ ergs.

The increased kinetic energy is due to an increase in the neutrino reheating behind  the shock, starting at about 200~ms postbounce.  This can be understood in terms of the location of the largest radius at which the MSW resonance energy falls below the electron-neutrino chemical potential.   As long as the resonance energy is below the chemical potential, there are many neutrinos available to participate in oscillations.  However, until about 200~ms postbounce, the last crossing is well below the neutrinosphere, so the formation of sterile neutrinos has little effect of the explosion dynamics.  

Beginning at about 200~ms, however, the last crossing has moved to just below the neutrinosphere.  In this case, the efficient transport of energetic neutrinos from the core to just below the neutrinosphere causes a heating of the surface of the protoneutron star and an enhanced luminosity of all three flavors of thermally produced neutrinos in addition to the transported flux.  This increases the neutrino luminosity at the neutrinosphere and ultimately energizes the shock as a fraction of these neutrinos deposit energy behind the shock.  

\begin{figure}[h!]
\centering
\includegraphics[width = 0.5\textwidth]{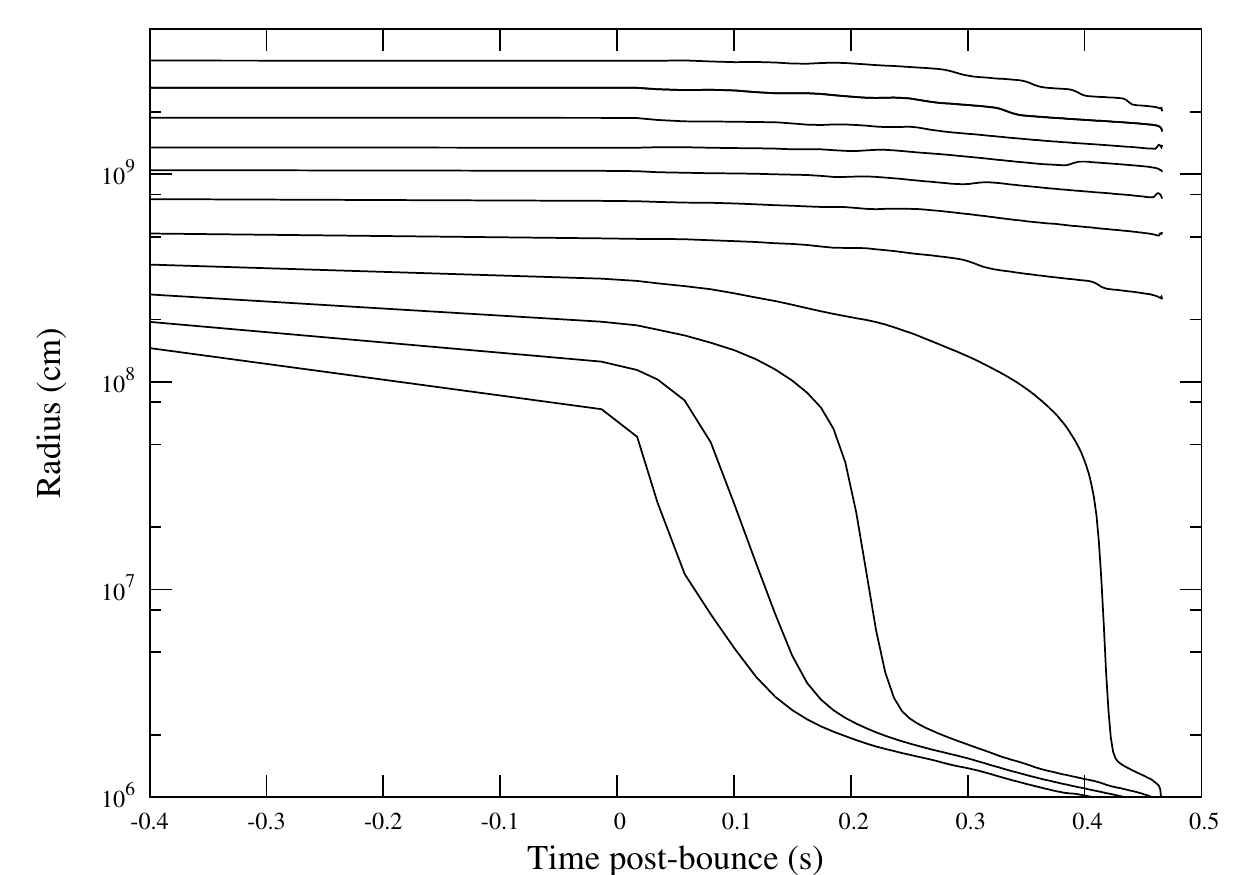}
\caption{Radius versus time postbounce for various mass elements in a simulation without convection and without sterile neutrinos.  Without additional neutrino reheating, the simulation collapses to a black hole at 0.4463~s postbounce.}
\label{fig:fig3}
\end{figure}

A key feature of the UND/LLNL supernova model is the inclusion of a dendritic neutron finger instability.  This is a doubly diffusive convective process that occurs near the surface of the nascent protoneutron star \cite{book,wilson1988}.   The presence of this instability enhances the neutrino luminosity by transporting lepton rich material to the protoneutron star surface.  The increased neutrino flux can lead to sufficient neutrino reheating of the stalled shock to cause a successful explosion.  However, the presence of this process in core-collapse supernovae is undecided and most core-collapse supernova models do not incorporate this effect \cite{bruenn2004}.   Hence, it is worthwhile to examine the robustness of the active-sterile mixing in a model which does not explode.

Figure~\ref{fig:fig3} shows the same 20~$M_{\odot}$ initial model but with all convection (including the outer dendritic convection) suppressed.  This results in a failed supernova that collapses to a black hole after about 0.45s.

\begin{figure}[h!]
\centering
\includegraphics[width = 0.5\textwidth]{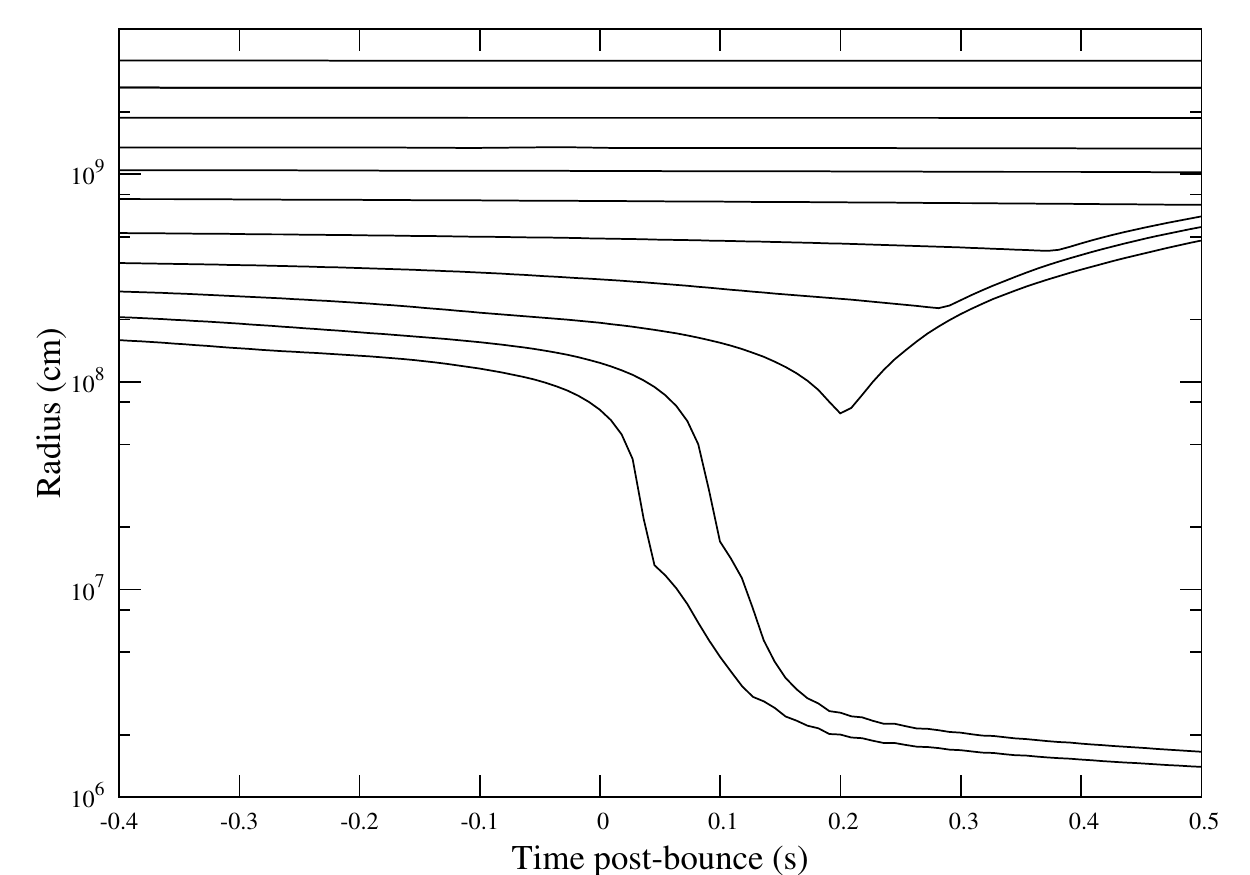}
\caption{Radius versus time postbounce for several mass elements in a simulation without convection, but including sterile neutrinos with $\Delta m_{s} = 5.012$~keV and  $\sin^{2} 2 \theta_{s} =  1.12\times 10^{-5}$.  The neutrino reheating provided by sterile neutrinos is sufficient to cause a successful explosion without convective processes present. }
\label{fig:fig4}
\end{figure}

Figure~\ref{fig:fig4}, however, shows the same initial model with suppressed convection, but in this case the active-sterile neutrino mixing is included with parameters  $\Delta m_{s} = 5.012$~keV and  $\sin^{2} 2 \theta_{s} =  1.12\times 10^{-5}$, consistent with the enhanced explosion energy of Fig.~\ref{fig:fig2}.  Note that in this case a robust explosion ensues, demonstrating that this effect can lead to explosions even for an otherwise nonexploding core-collapse model.  

Given the robustness of this explosion, we expect that a sterile neutrino can also enhance the explosion in other nonexploding one-dimensional models as well as nonexploding two- or three- dimensional models. This is because the transport of heating by sterile neutrinos is largely a surface phenomenon and is independent of 2- or 3D effects below (or above) the surface.

\subsubsection{Neutrino luminosity and flux}
\begin{figure}
        \centering
                    \includegraphics[width=0.5\textwidth]{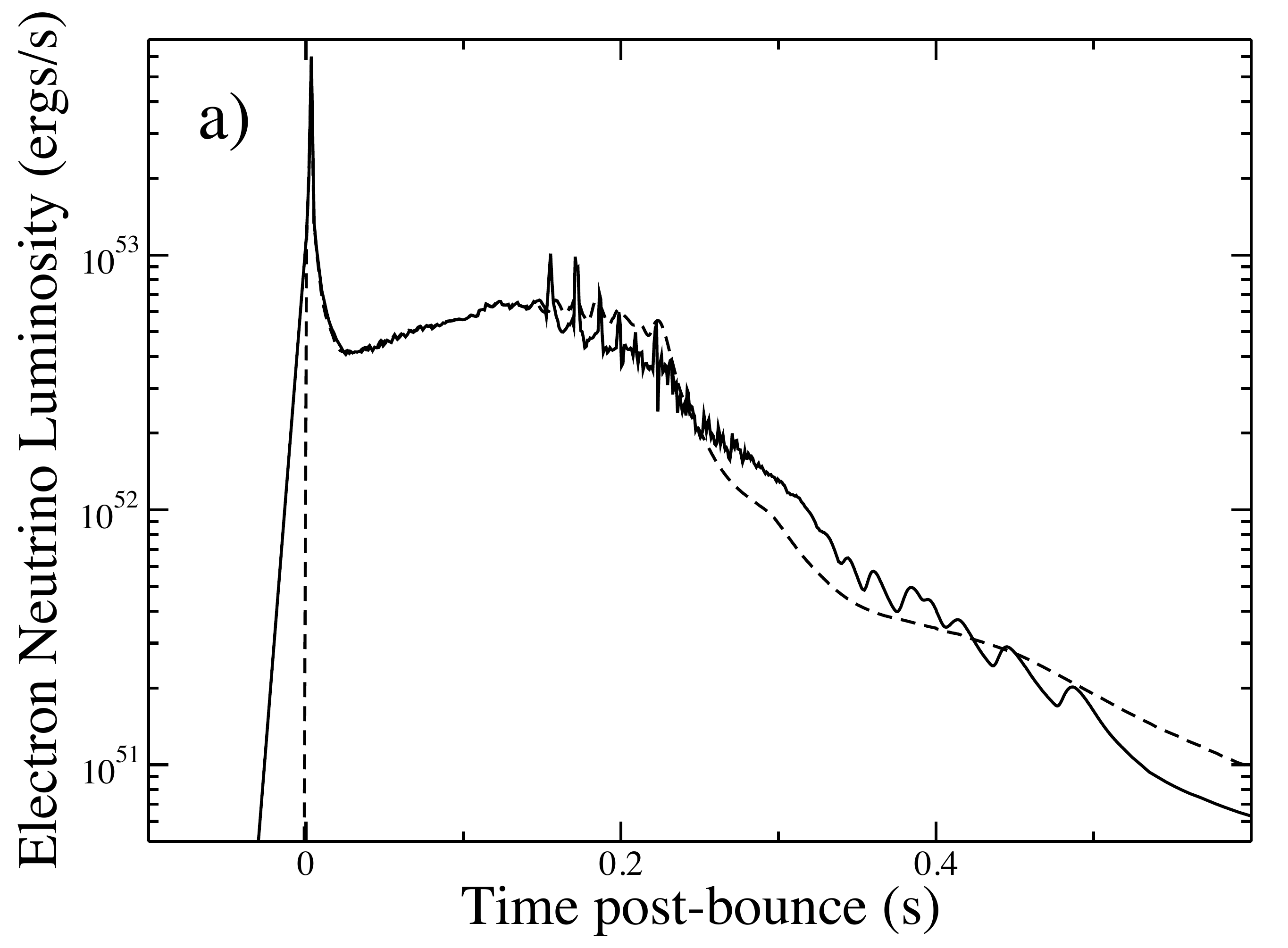}
                \includegraphics[width=0.5\textwidth]{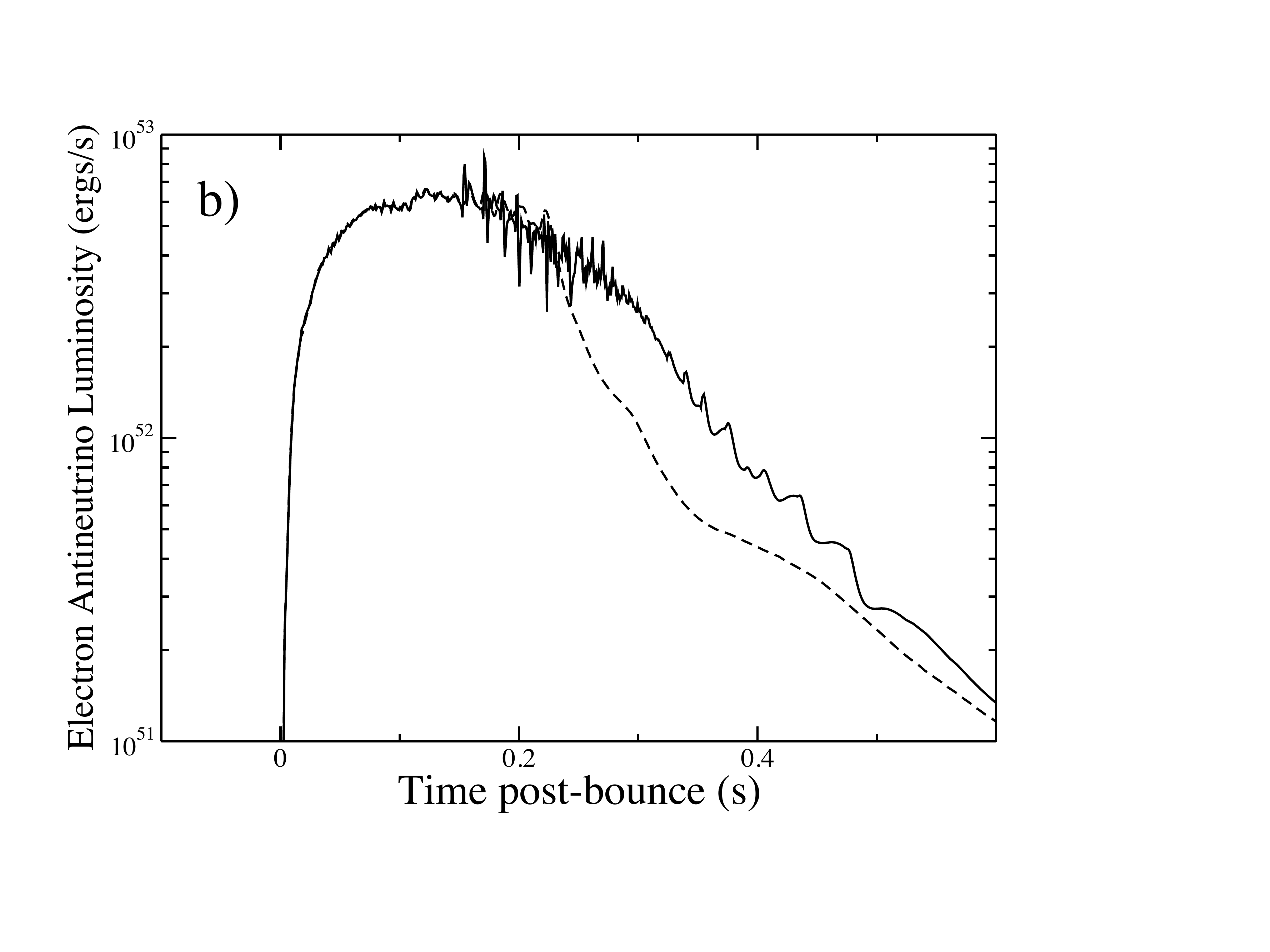}
                 \includegraphics[width=0.5\textwidth]{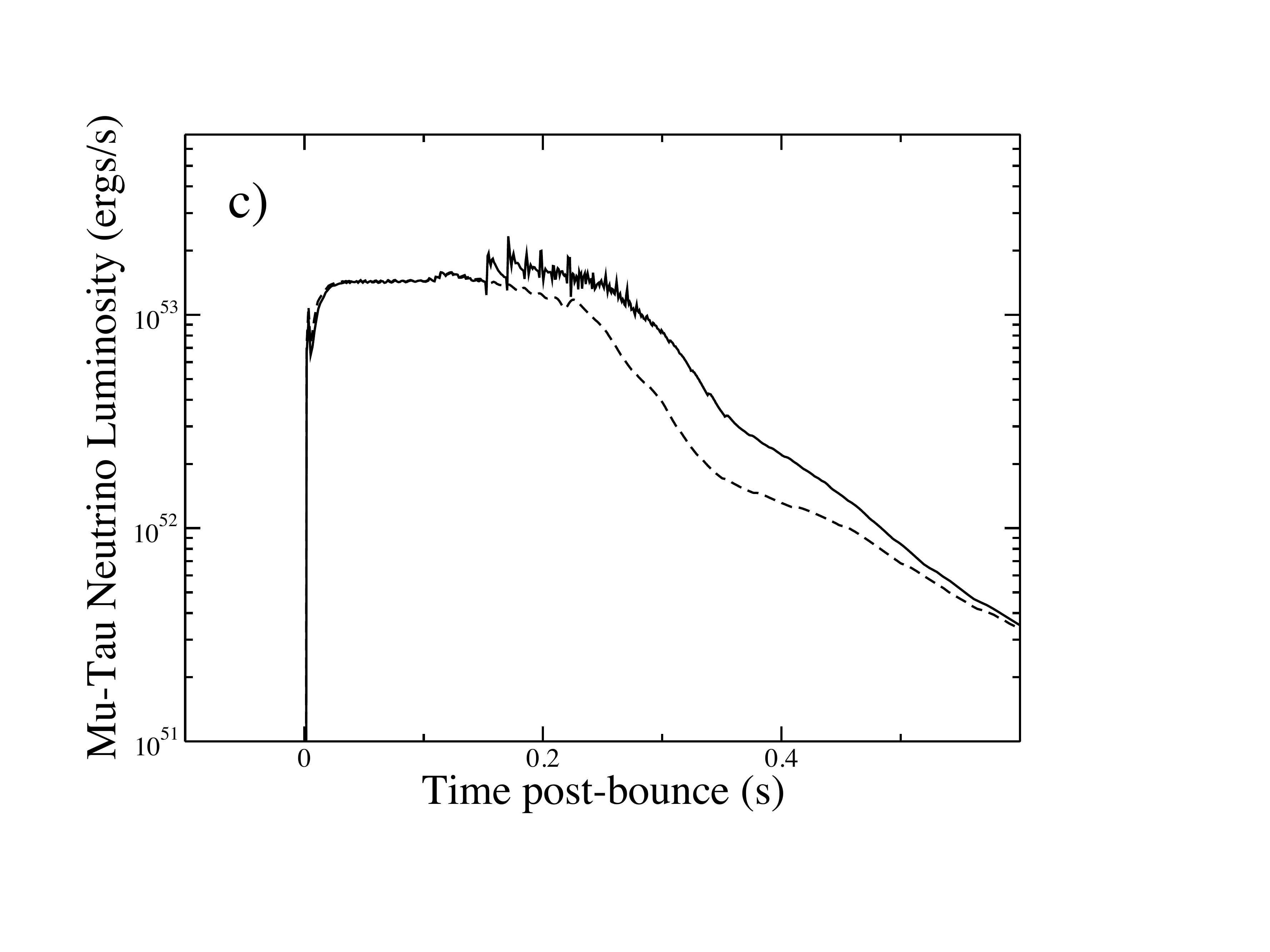}
         \caption{Observable neutrino luminosities versus time postbounce. (a) is for the electron neutrinos, (b) is for the electron antineutrinos, and (c) is for the $\mu,\tau$ neutrinos.  The solid line is with a 5.012~keV sterile neutrino with $\sin^{2} 2 \theta_{s} = 1.12 \times 10^{-5}$ and the dashed line is without a sterile neutrino present. }
        \label{fig:fig5}
\end{figure}

\begin{figure}
        \centering
                \includegraphics[width=0.5\textwidth]{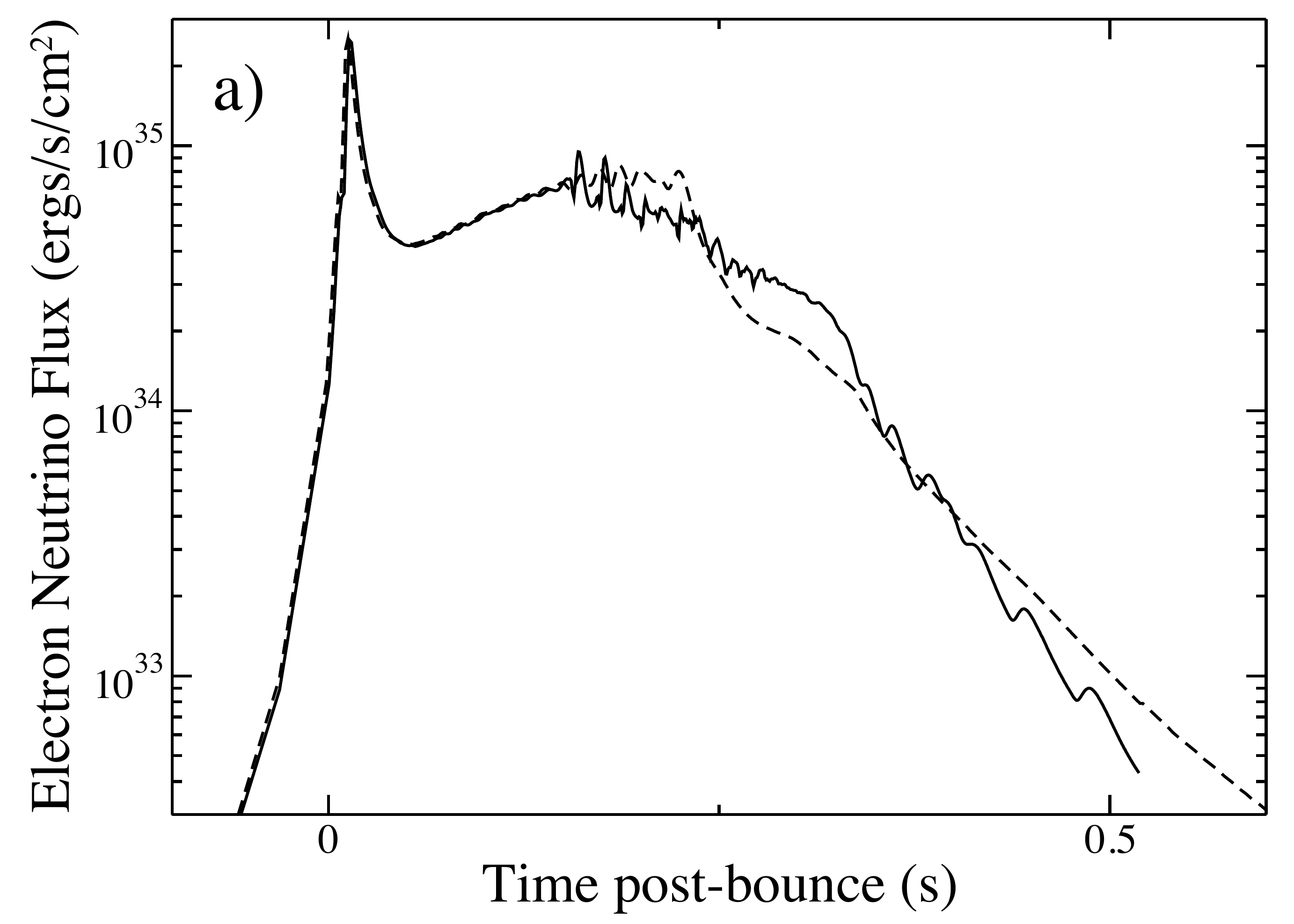}
                  \includegraphics[width=0.5\textwidth]{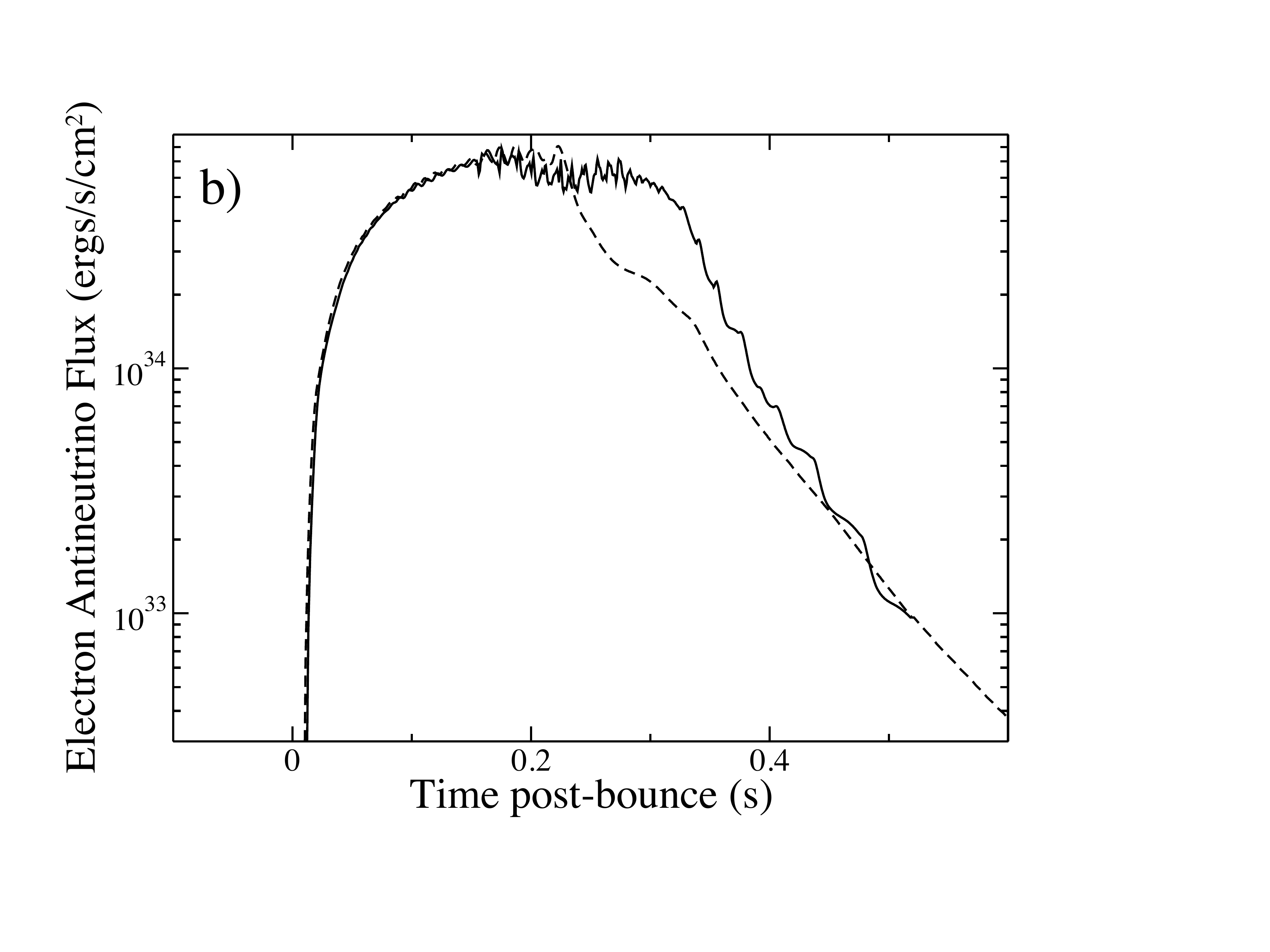}
                \includegraphics[width=0.5\textwidth]{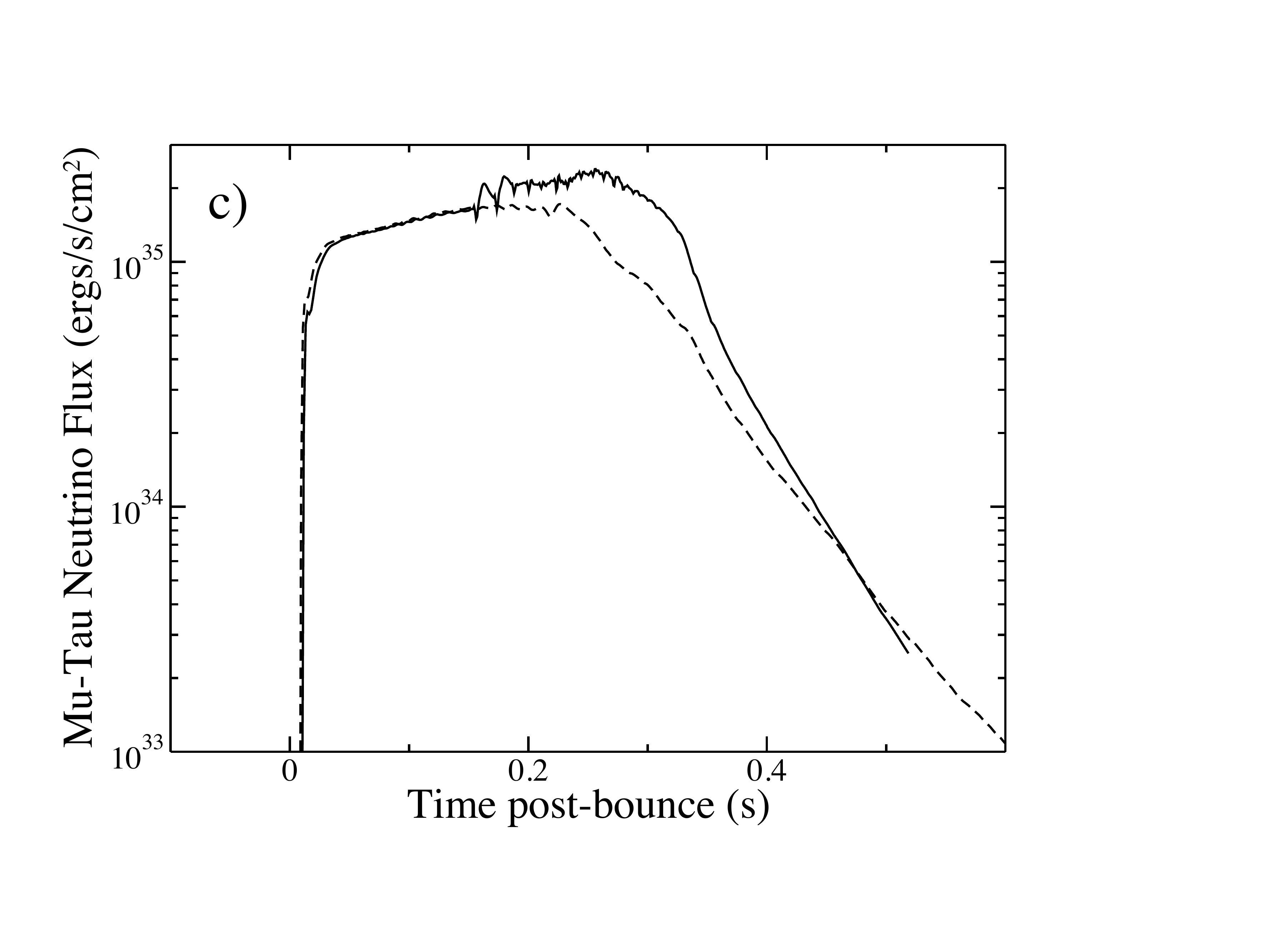}
        \caption{Neutrino fluxes at the neutrinosphere versus time postbounce. (a) is for the electron neutrinos, (b) is for the electron antineutrinos, and (c) is for the $\mu,\tau$ neutrinos. The solid line is a model including  oscillations with a 5.012~keV sterile neutrino with $\sin^{2} 2 \theta_{s} = 1.12 \times 10^{-5}$  and the dashed line is for a model without a sterile neutrino. }
        \label{fig:fig6}
\end{figure}

Figures \ref{fig:fig5} and  \ref{fig:fig6} show the observable luminosity and flux, respectively, versus time postbounce for all three neutrino flavors.  From a comparison of these two figures, one can get some insight as to what is going on.  One striking feature is the appearance of episodic neutrino bursts  with a period of $30-40$ ms in the luminosity and flux in the model with a sterile neutrino.  This is present in simulations both with and without the neutron finger instability.  A similar feature was seen in Ref.~\cite{esmaili2014}, but for a very low mass sterile neutrino $\sim 3-6$~eV. 

This episodic variation of the neutrino luminosity is easy to understand.  The neutrino photospheric luminosity of the protoneutron star  is fixed  by the ratio of the total internal energy to the neutrino  diffusion time.  When the neutrino chemical potential falls above the resonance energy, those neutrinos in the energy groups corresponding to the resonance energy and width immediately have a diffusion time scale drastically shortened by the free streaming of the sterile neutrinos to just below the neutrinosphere.  Hence, there is a precipitous drop in the average diffusion time scale and a concomitant increase in the luminosity.  

Once this burst of luminosity depletes the neutrinos at the neutrinosphere, a significant fraction of available (mostly antielectron) neutrinos is depleted, the process is shut off and the luminosity actually decreases (relative to a model with no sterile neutrino) until neutrinos can diffuse back into the depleted energy and spatial groups at the neutrinosphere.   Also, the forward scattering potential [Eq.~(\ref{eq:pot})] shifts as the interior electron antineutrino number density is depleted.  The resonance energy [Eq.~(\ref{eq:res})] also shifts, contributing slightly to the cycle. 

\begin{figure}[h]
   \centering
   \includegraphics[width = 0.5\textwidth]{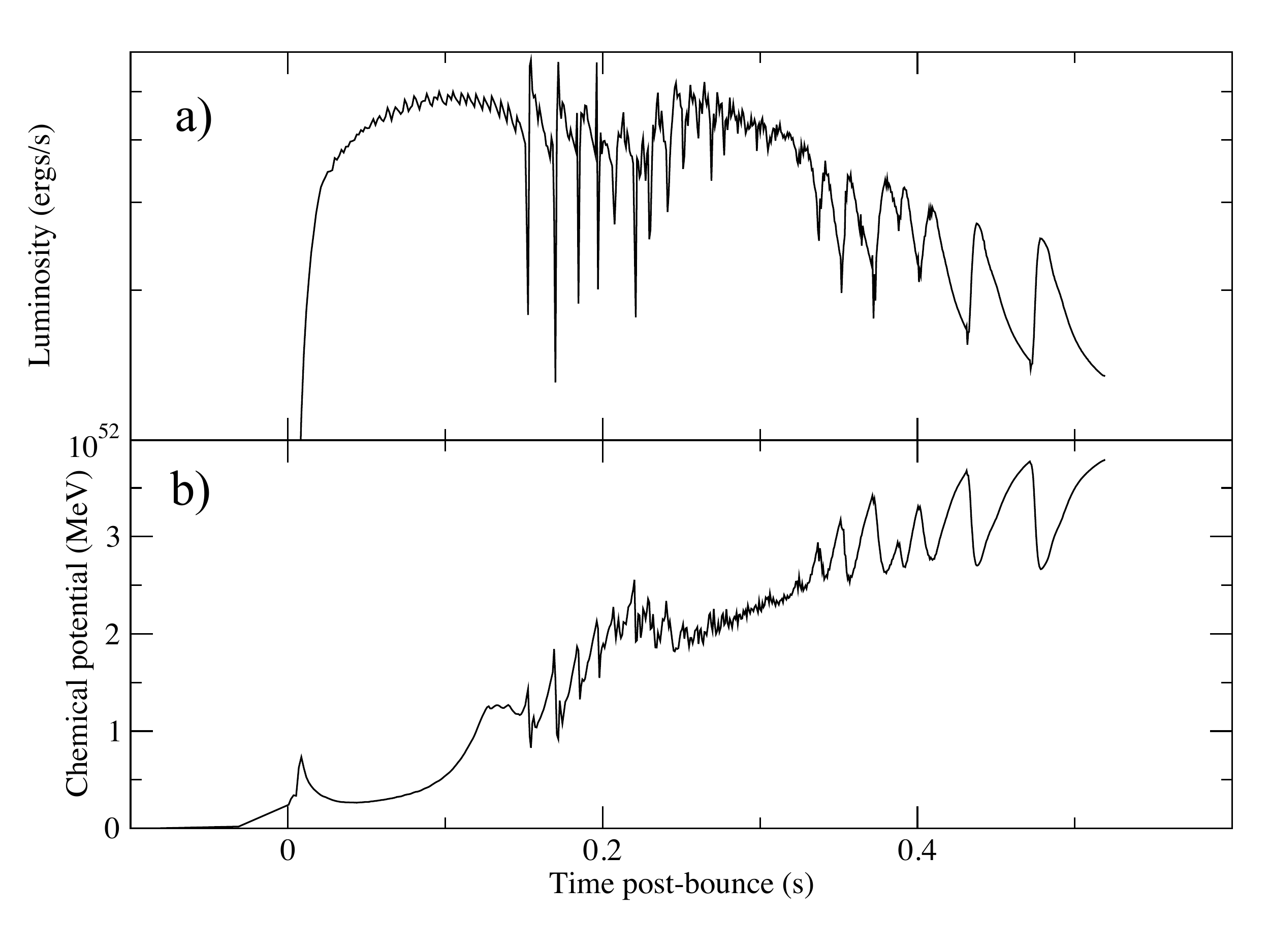} % requires the graphicx package
   \caption{(a) Electron antineutrino luminosity versus time at a radius of $\sim20$ km.  (b) Neutrino chemical potential vs time near the neutrinosphere. }
   \label{fig:fig7}
\end{figure}

Figure \ref{fig:fig7} illustrates this interpretation.  This figure shows  the electron antineutrino luminosity and chemical potential vs~time postbounce near the neutrinosphere.  Note that the chemical potential and luminosity are anticorrelated.  After bounce, the neutrino chemical potential and density  increase until the chemical potential  exceeds the resonance energy.
Above that threshold there is a spike in the luminosity as active neutrinos are converted to sterile neutrinos and transported out from the interior. As the chemical potential and density diminish, however, the luminosity also drops by the depletion of available electron antineutrinos, shutting off the process until neutrinos can diffuse back into the depleted region.

Comparing the top panels in Fig.~\ref{fig:fig5} and  \ref{fig:fig6}, one can see that the electron-neutrino luminosity increases after 200 ms, while the flux decreases.  This  enhanced luminosity  continues for about 500~ms into  the  supernova explosion.  This is just the critical time for enhancing the delayed heating mechanism~\cite{wilson2005,book}. 
This increased luminosity results from the sterile neutrino to active conversion below the neutrinosphere.  This leads to  heating and extending the radius of the neutrinosphere.  

\begin{figure}[t]
   \centering
   \includegraphics[width = 0.5\textwidth]{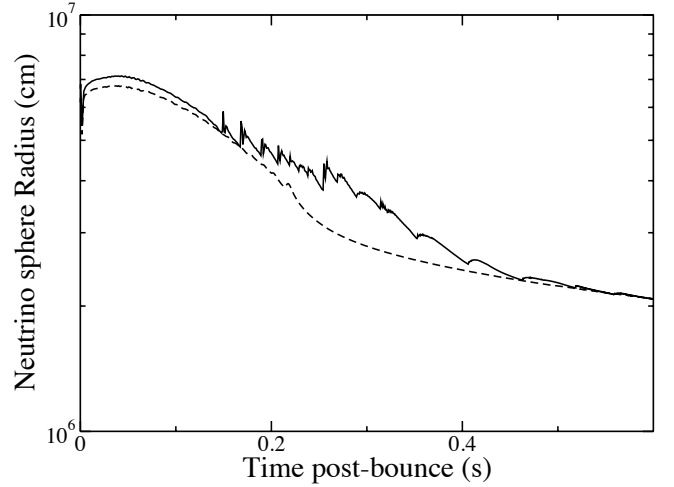} % requires the graphicx package
   \caption{Location of the neutrinos sphere vs time for the first 500~ms postbounce.  The solid line is for a sterile neutrino with $\Delta m_{s} = 5.012$ keV and  $\sin^{2} 2 \theta_{s} =  1.12\times 10^{-5}$ and the dashed line is without a sterile neutrino.  Note that the sawtooth effect in the neutrinosphere radius is a graphic artifact of choosing a single zone as the neutrinosphere when in fact it is spread over several zones. }
   \label{fig:fig8}
\end{figure}

Figure \ref{fig:fig8} shows the neutrinosphere radius vs. time from models with and without a sterile neutrino in the simulation.  
Figure \ref{fig:fig8}  shows that in models with a sterile neutrino, the radius extends outward by up to about a factor of 1.4.  This factor in itself would lead to  a factor of 2 increase in luminosity.

The depletion in the flux, however, in Fig.~\ref{fig:fig6}(a) is due  to the fact that electron neutrinos are suppressed by the oscillations with the sterile neutrino, as discussed below.  Similarly, although the average electron antineutrino flux is moderately enhanced ($\sim 20$\%) at the neutrinosphere, the luminosity is increased by a factor of 3 after about 200~ms due to the extended radius of the neutrinosphere.  Finally, although the $\nu_\mu$ and $\nu_\tau$ flux is slightly increased by the thermal production from heating near the neutrinosphere, the luminosity increases by a factor of 3 at around 300 ms due to the extension of the neutrinosphere.

To better clarify this, we now analyze the interior structure of the exploding star  in detail at three time slices, i.e.~$t = 0$ (core bounce), 270~ms, and 520~ms postbounce.

\subsubsection{Core bounce}

\begin{figure}[b]
   \centering
   \includegraphics[width =0.5 \textwidth]{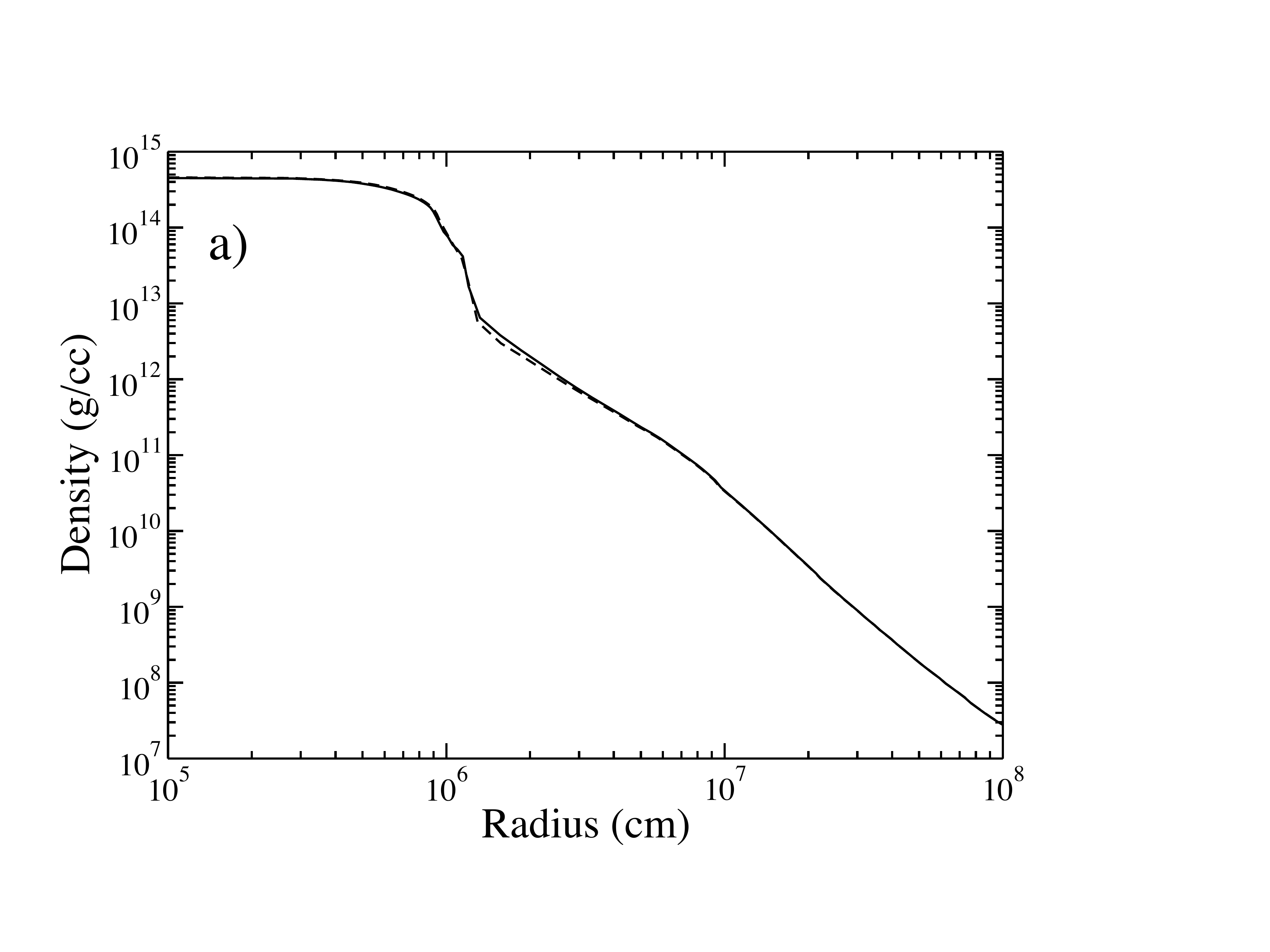} % requires the graphicx package
   \includegraphics[width = 0.5\textwidth]{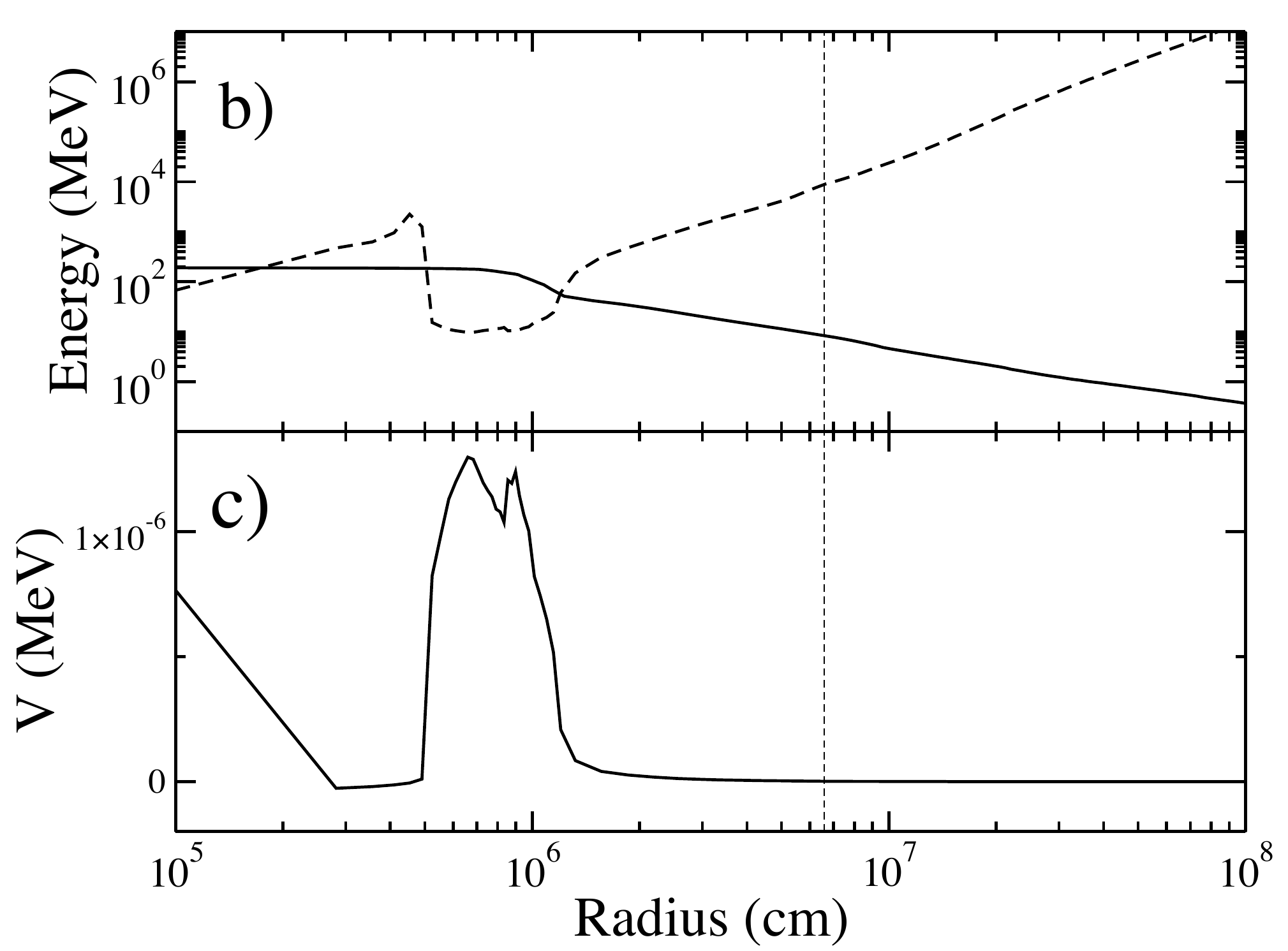} % requires the graphicx package
   \caption{Upper panel (a) shows the density profile at core bounce.  The solid line shows a model in which oscillations between  sterile neutrinos and $\nu_e$s or $\bar \nu_e$s are allowed and the dashed line is for a model without a sterile neutrino.   The lower panel (b) shows the neutrino chemical potential (solid line) and resonance energy (dashed line) versus radius  and (c) shows the neutrino forward scattering potential vs radius at bounce in the SN simulation.  The vertical dashed line marks the location of the neutrinosphere. }
   \label{fig:fig9}
\end{figure}

Figure \ref{fig:fig9} provides a snapshot of the protoneutron star right at bounce ($t = 0$).  As one can see from Fig.~\ref{fig:fig9}(a), the density profile is extended as material is still falling onto the protoneutron star.   Hence, the  neutrinosphere of the forming neutron star is at about 25~km.  At this time a shock is just beginning to form at about $10^{6}$ cm.   

Figure~\ref{fig:fig9}(b) shows the resonance energy and  neutrino chemical potential and the forward scattering potential versus radius at the time of bounce.  Figure \ref{fig:fig9}(b) shows that, although  resonant active-sterile neutrino oscillations have begun  at the time of core bounce, the oscillations occur in two  narrow regions  deep  inside of the protoneutron star.  These regions correspond to the location of the shock.  The density fluctuation at the location of the shock causes the resonance energy to fall below the chemical potential and a dip in the forward scattering potential.   The regions of lower resonance energy at the time of bounce are large enough to allow a significant number of neutrinos to oscillate from electron neutrinos to sterile neutrinos.  However, at a radius of about 15~km, the resonance energy becomes too high for resonant oscillations to be important.  The plot of the forward scattering potential in Figure \ref{fig:fig9}(b) shows that the scattering potential is positive everywhere, and thus the oscillations occur entirely for the electron neutrino  and not for its antineutrino.  This changes at later times, however.

\begin{figure}[h]
   \centering
   \includegraphics[width = 0.5\textwidth]{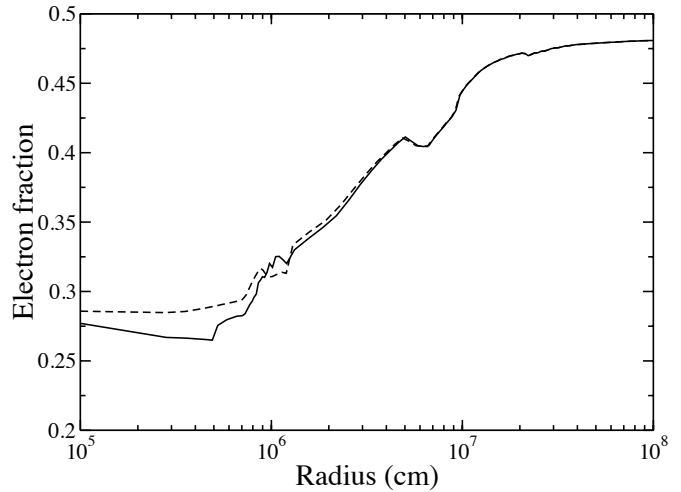} % requires the graphicx package
   \caption{Electron fraction versus radius at bounce.  The solid line is for a sterile neutrino with $\Delta m_{s} = 5.012$~keV and  $\sin^{2} 2 \theta_{s} =  1.12\times 10^{-5}$ and the dashed line is without a sterile neutrino.}
   \label{fig:fig10}
\end{figure}

Although at this point it is too early for the sterile neutrino oscillations to have significantly affected the evolution, there is already some evidence that these oscillations lead to feedback effects on the local environment.  Figure~\ref{fig:fig10} shows the electron fraction versus radius at bounce.  The electron fraction in the protoneutron star core is depleted when a sterile neutrino is allowed in the simulation.  The oscillation of electron neutrinos to sterile neutrinos inside the protoneutron star frees up neutrino phase space and increases the rate of electron capture on protons.  This decreases the electron fraction in the regions where electron neutrinos are oscillating to sterile neutrinos.  The electron fraction in the protoneutron star core remains depleted throughout the remainder of the supernova evolution due to this effect.

Conversely, at radii $\sim10$ to 15~km, the opposite effect begins to occur.  Sterile neutrinos that were created in the protoneutron star core resonate back to electron neutrinos in this region and lead to slightly increased electron neutrino captures on neutrons.  This results in an increase in the electron fraction in this region compared to a  simulation without a sterile neutrino present, as can be seen in Fig.~\ref{fig:fig10}.  This effect continues for some time after bounce and results in a peak in the electron fraction just above 10~km.  Later, this region of increased electron fraction will begin to diminish through electron captures and will result in an increased emission of electron neutrinos just below the neutrinosphere.  Nonetheless, at the time of bounce, the net effect in this region remains -- i.e., a slight neutrino cooling occurs due to the emission of electron neutrinos -- despite the deposition of neutrinos from sterile neutrino oscillations.   It is not until later times that the net neutrino heating becomes enhanced.

\subsubsection{270~ms postbounce}

By $t = 270$~ms postbounce, the resonance energy has fallen below the neutrino chemical potential almost everywhere interior to the neutrinosphere of  the protoneutron star.  However, it is important to note that the resonant oscillations occur primarily for antineutrinos.  The combination of high core densities and low electron fractions effectively ``shut off" the resonant electron-neutrino oscillations inside of the protoneutron star.  The high density suppresses coherent oscillations.  Also, the resonance energy, which is inversely related to the density, is driven to very low energies.  In addition, the electron fraction falls below one third,  which results in a  negative forward scattering potential [see Eq.~(\ref{eq:potential})].  Thus, any resonance occurs for electron antineutrinos, rather than electron neutrinos. 

\begin{figure}[t]
   \centering
\includegraphics[width =0.5\textwidth]{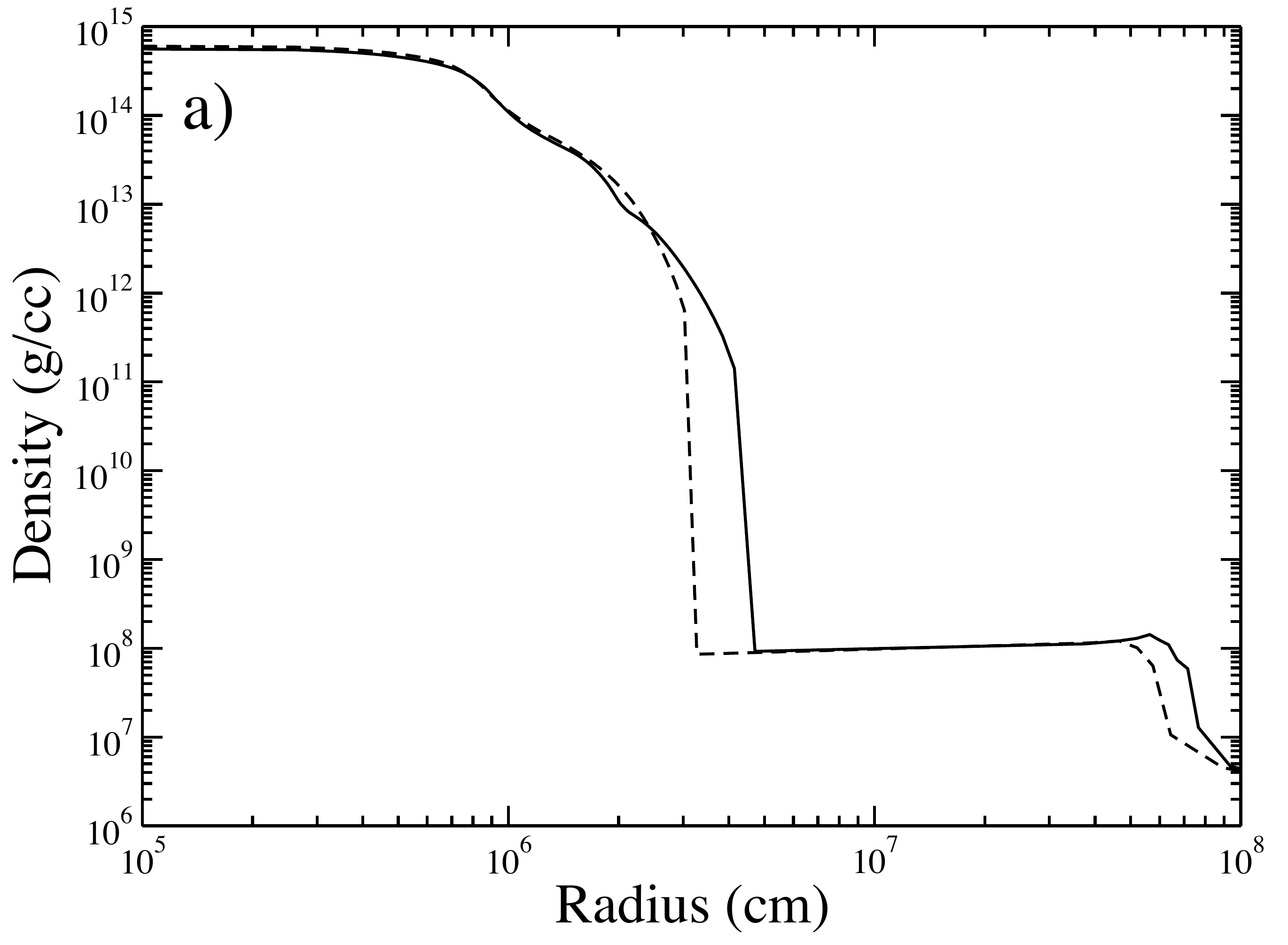}
   \includegraphics[width = 0.5\textwidth]{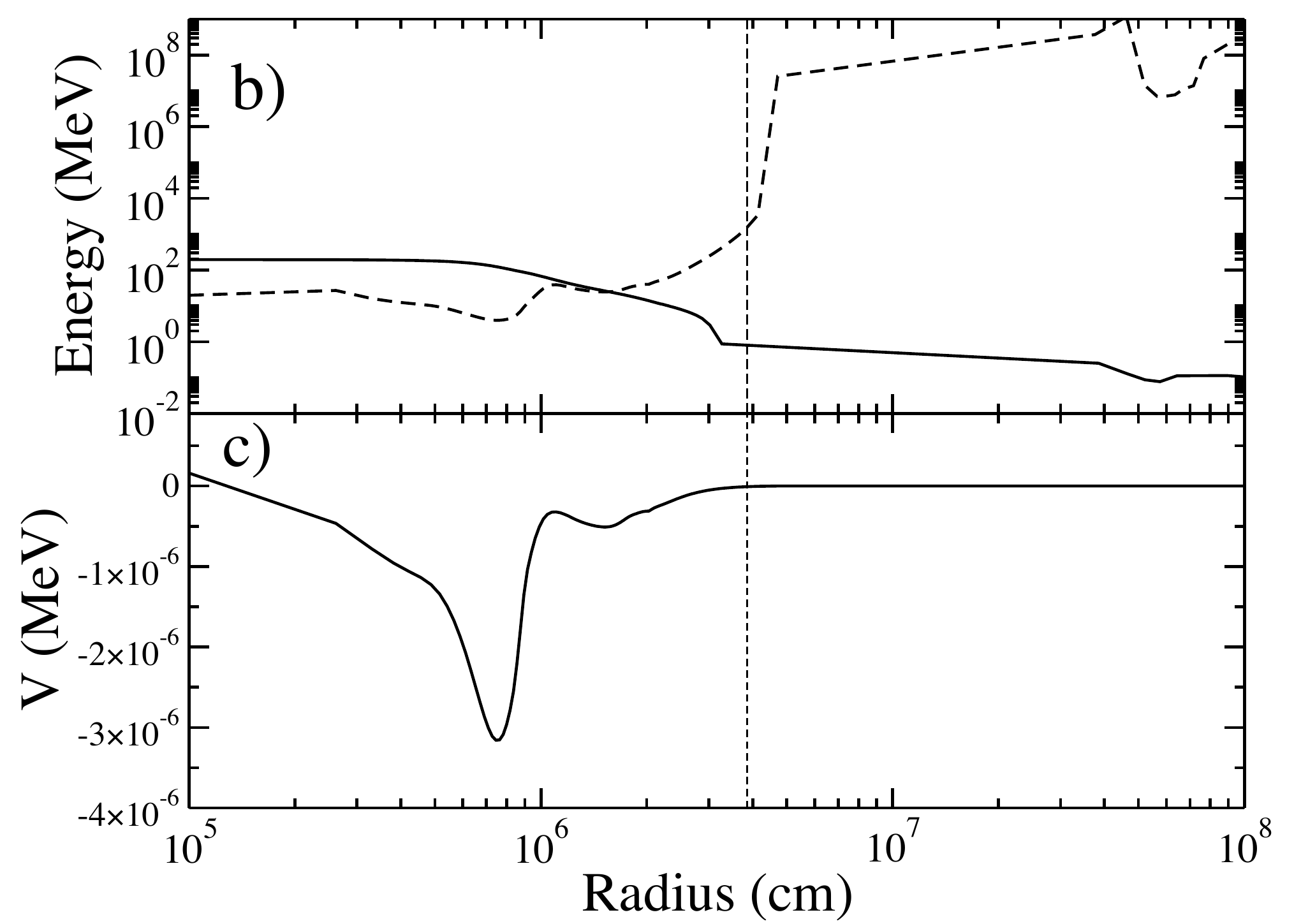} % requires the graphicx package
   \caption{Upper panel (a) shows density versus radius at 270~ms postbounce in the SN simulation.   The solid line is for a sterile neutrino with $\Delta m_{s} = 5.012$ keV and  $\sin^{2} 2 \theta_{s} =  1.12\times 10^{-5}$ and the dashed line is without a sterile neutrino.  Lower panel (b) shows the neutrino chemical potential (solid line) and resonance energy (dashed line) versus radius and (c) shows the neutrino forward scattering potential vs radius.  The vertical dashed line marks the location of the neutrinosphere in the model with a sterile neutrino.}
   \label{fig:fig11}
\end{figure}

Figure \ref{fig:fig11} shows the density structure, neutrino chemical potential, resonance energy, and forward scattering potential  at 270~ms  postbounce.  There is an inflection in the density and a corresponding dip in the forward scattering potential and resonance energy that is caused by the heating of material after  the conversion of sterile neutrinos back to active neutrinos at this location.  Here, one can see that by this time the resonance energy falls below  the neutrino chemical potential until about 2 km below the neutrinosphere.  Except for a small region in the very center of the protoneutron star, the forward scattering potential is negative.  Thus, the resonant oscillations occur almost exclusively for electron antineutrinos.  
The electron antineutrinos resonantly oscillate to sterile antineutrinos essentially through the entire region where the resonance energy is below the neutrino chemical potential.  This enhances the cooling of the protoneutron star.   The sterile antineutrinos produced in the protoneutron star core oscillate back to electron antineutrinos in a  region ranging from $\sim$ 2 to 2.3 km  below the neutrinosphere. 

\begin{figure}[b]
   \centering
   \includegraphics[width = 0.5\textwidth]{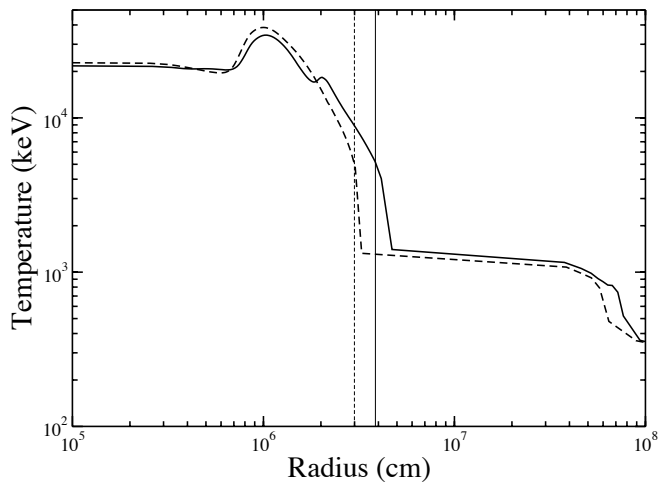} % requires the graphicx package
   \caption{Temperature versus radius at 270~ms postbounce. The solid line is for a sterile neutrino with $\Delta m_{s} = 5.012$~keV and  $\sin^{2} 2 \theta_{s} =  1.12\times 10^{-5}$ and the dashed line is without a sterile neutrino. The dashed vertical line shows the neutrinosphere in a model without a sterile neutrino.  The vertical solid line shows the expanded neutrinosphere in the model with a sterile neutrino.}
   \label{fig:fig12}
\end{figure}

The oscillations between electron antineutrinos and sterile antineutrinos are reflected in the density and temperature profiles on Figs.~\ref{fig:fig11}(a) and \ref{fig:fig12}.  The neutrino losses from the protoneutron star core and the additional neutrino energy deposited just below the neutrinosphere expand the surface of the  protoneutron star, as reflected in the density profile in the upper panel of  Figure \ref{fig:fig11}(b).   

Figure \ref{fig:fig12} similarly shows the temperature versus radius profile at 270~ms postbounce.  The interior temperature in the protoneutron star core is slightly decreased due to the energy lost to sterile neutrino oscillations.  A more dramatic effect, however, is that up to about 20~km, the sterile antineutrinos can oscillate back to electron antineutrinos.  This enhanced flux of energetic active antineutrinos increases the  heating  near the neutrinosphere, thereby causing the star to expand near the surface.

The extended density profile means that the neutrinos decouple at larger radii.  However, due to the increased heating near and below the neutrinosphere, the temperature at the neutrinosphere is unchanged for both simulations, as shown by the vertical lines in Fig.~\ref{fig:fig12}.  Thus, the emergent neutrinos have roughly the same average energy in both simulations, but the larger radius of the neutrinosphere results in increased luminosities (by a simple factor of $r^2$) for all three neutrino flavors. The enhanced neutrino heating behind the shock resulting from the enhanced neutrino luminosity  is responsible for the increased kinetic energy in the explosion, as shown in Fig.~\ref{fig:fig2}.

\subsubsection{520~ms postbounce}

\begin{figure}[t]
   \centering
   \includegraphics[width =0.5 \textwidth]{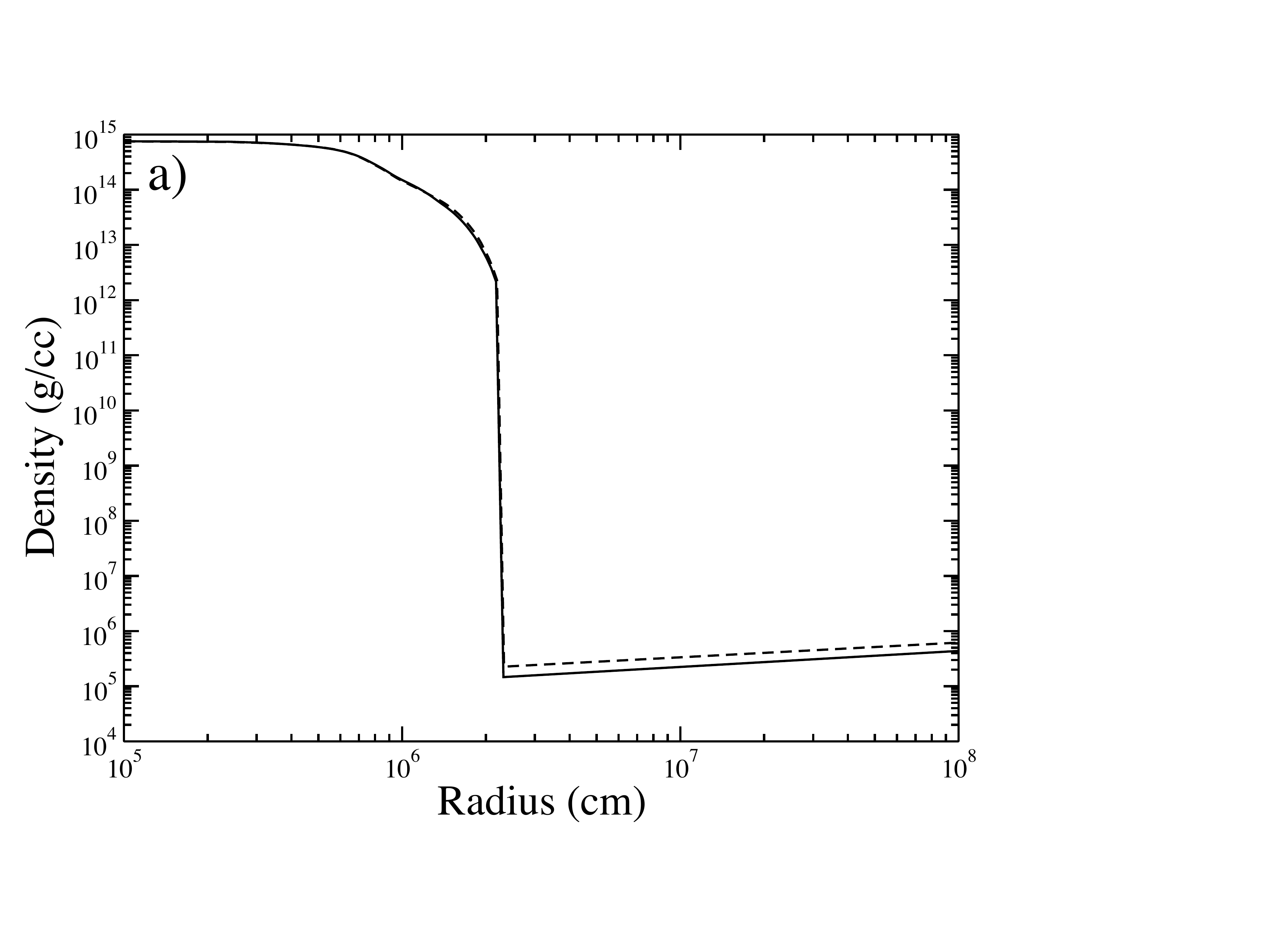} % requires the graphicx package
   \includegraphics[width = 0.5\textwidth]{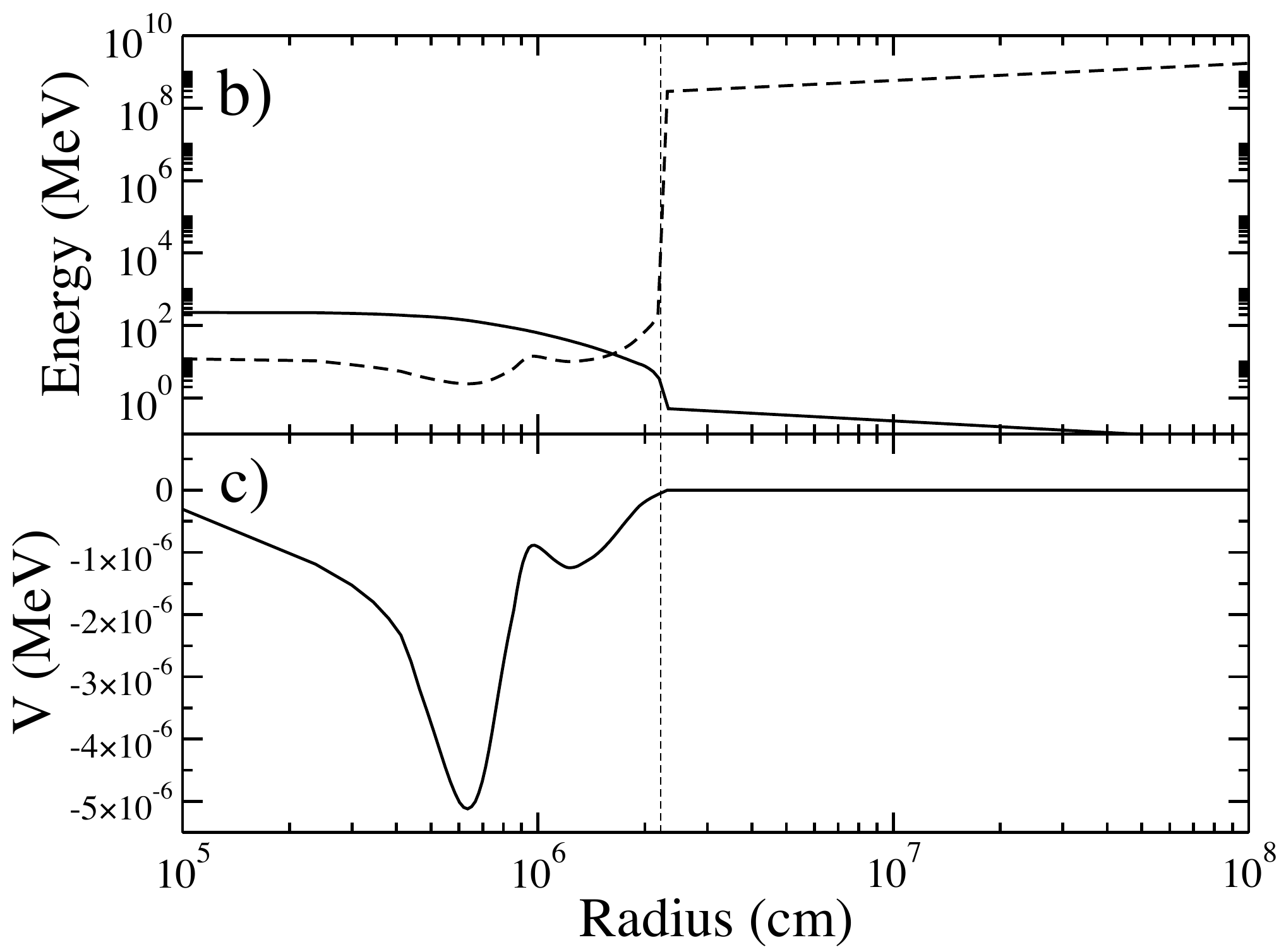} % requires the graphicx package
    \caption{Upper panel (a) shows the density profile versus radius at 520 ms postbounce for a simulation without a sterile neutrino (dashed line) and with a sterile neutrino (solid line) with $\Delta m_{s} = 5.012$ keV and  $\sin^{2} 2 \theta_{s} =  1.12\times 10^{-5}$.  The lower panel (b) shows the neutrino chemical potential (solid line) and resonance energy (dashed line) versus radius and (c) shows the neutrino forward scattering potential vs. radius at 520 ms postbounce  in the SN simulation.  The vertical dashed line marks the location of the neutrinosphere. }
   \label{fig:fig13}
\end{figure}

Figure \ref{fig:fig13}(a) shows the density profile and resonance parameters at 520~ms postbounce.  One can see that by this time, the heating and expansion of the neutrinosphere  has nearly subsided. This is because the cooling by neutrino emission at the neutrinosphere exceeds the heating by sterile neutrino conversion to energetic active neutrinos.   Hence, the protoneutron star  settles to the same radial density and temperature profiles as for the simulation without a sterile neutrino.  Similarly, as shown in Fig.~\ref{fig:fig5} the neutrino luminosities begin to converge to those without a sterile neutrino.
 
 As can be seen in the top  panel of Fig.~\ref{fig:fig13}(b), the resonance energy is below the neutrino chemical potential throughout most of the interior of the protoneutron star.   Hence, the  sterile neutrino oscillation condition is satisfied nearly all the way to the neutrinosphere.  As a result, there is little added heating and expansion near the surface.  Also note that the forward scattering potential, shown in the lower panel of Fig.~\ref{fig:fig13}b, is negative everywhere inside of the protoneutron star.  Thus, the mixing only occurs for antineutrinos.   
 
\begin{figure}[h]
   \centering
   \includegraphics[width = 0.5\textwidth]{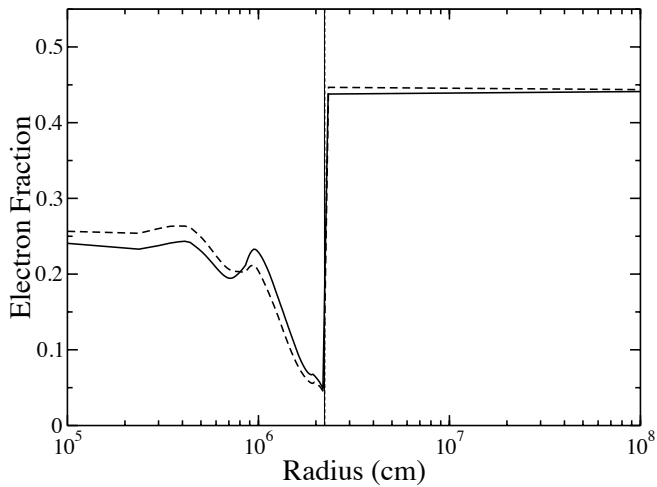} % requires the graphicx package
   \caption{Electron fraction versus radius at 520 ms postbounce.  The solid line is for a sterile neutrino with $\Delta m_{s} = 5.012$ keV and  $\sin^{2} 2 \theta_{s} =  1.12\times 10^{-5}$ and the dashed line is without a sterile neutrino. Electron fraction is depleted in the inner core, indicative of sterile neutrino oscillations.}
   \label{fig:fig14}
\end{figure}

The antineutrino oscillations also alter the matter composition and cause feedback effects on the environment.  Figure \ref{fig:fig14} shows the electron fraction versus radius at 520~ms postbounce.  The decreased electron fraction $Y_e$ in the protoneutron star is a relic of the electron neutrino to sterile neutrino mixing that began around the time of bounce.  The electron fraction is then slightly enhanced just below the neutrinosphere due to the antineutrino oscillations that dominate after bounce.

\subsubsection{Observable effects}
 
One should, of course, ask whether it is possible to determine that such oscillations indeed have occurred in an observed supernova.  It is important to note that the efficient cooling of the protoneutron star via electron antineutrino oscillations not only enhances the kinetic energy of the explosion, but also increases the energy released in the form of neutrinos.  As we showed in Fig.~\ref{fig:fig5}(a), oscillations with a sterile neutrino have little effect on the electron-neutrino luminosity.  Although, after 300~ms, by which time the resonance energy has fallen below the chemical potential, the electron neutrino luminosity is slightly increased compared to the simulation without a sterile neutrino.  

As noted above, the enhanced luminosity is mainly due to the expansion of the neutrinosphere.    The electron antineutrino luminosity, however, as shown in Fig.~\ref{fig:fig5}(b), exhibits  a more pronounced  enhancement in luminosity (a factor of 2) after 300~ms. Unfortunately, the neutrino luminosities from the simulations with and without a sterile neutrino begin to converge at  times later than about 500~ms.   Nevertheless, one can speculate that a future very well time resolved neutrino light curve  from the first 500~ms, might indicate an enhanced luminosity in electron antineutrinos.

Although the $\mu-\tau$ neutrinos are not directly involved in the sterile neutrino oscillations, their luminosity is altered with the inclusion of a sterile neutrino.  The active-sterile neutrino oscillations cause heating below the neutrinosphere, so there is enhanced thermal production of $\mu-\tau$ neutrinos and they  decouple at a larger radius.  This results in the enhanced neutrino luminosity shown in Fig.~\ref{fig:fig5}(c).  The $\mu-\tau$ neutrinos do not experience the same losses due to sterile neutrinos and only exhibit  the effects of the increased heating and expansion of the radius of the neutrinosphere.  Thus, a future detector may also detect an enhancement in $\mu-\tau$ neutrinos.

One can also speculate that a well time resolved future supernova event might even exhibit the periodic bursts associated with the cycle of diminished diffusion time followed by depletion of available neutrinos.  This would be a definitive signature of the process discussed here, although one should caution that convection in 2D or 3D may diminish the periodic burst effect as the neutrino sphere becomes more diffuse.

\section{Conclusion \label{sec:conc}}

We have shown that  the inclusion of $\nu_{e} \xleftrightharpoons[\hspace{0.25cm}]{} \nu_{s}$ mixing can result  in alterations to the observable neutrino luminosities and an increase in the shock energy for a broad range of mixing angles and sterile neutrino masses.  Indeed, the region of largest enhancement is consistent with most cosmological constraints on dark matter candidates.  

The heating and cooling via electron antineutrino to sterile neutrino oscillations alters the emission of all three neutrino flavors, leading  to increased neutrino  luminosities emerging from the neutrinosphere at early times.  
It is important to note that the efficient cooling of the protoneutron star via electron antineutrino oscillations enhances the kinetic energy, but the cooling also increases the energy released in the form of neutrinos.  Hence, one may find a signature of this process in the enhanced early luminosity  of electron antineutrinos and $\mu,\tau$ neutrinos. 

One can also speculate that a well time resolved future supernova neutrino light curve detection might find evidence of the periodic bursts associated with the cycle of diminished diffusion time followed by depletion of available neutrinos.  This would be a definitive signature of the process discussed here, although one should caution that convection in 3D may diminish the periodic burst effect.

We analyzed the enhancement to the shock energy for a representative sterile neutrino with $\Delta m_{s} = 5.012$~keV and  $\sin^{2} 2 \theta_{s} =  1.12\times 10^{-5}$ and  found  that the enhanced kinetic energy is primarily due to increased neutrino luminosities from the neutrinosphere.  This results from a combination of a shorter diffusion time and an extended density  profile of the protoneutron star so that  neutrinos decouple at a larger radius.  This latter effect  results in larger thermally produced luminosities for all three neutrino flavors emerging from the neutrinosphere.  Ultimately, antineutrino oscillations dominate over neutrino oscillations due to the low electron fraction in the protoneutron star core. 

As a final caveat, however, we note that  simulations described  here considered only coherent, adiabatic transformations between electron and sterile neutrinos.  Additional oscillations deep in the core may occur due to scattering-induced decoherence, which may increase the luminosity of the high energy electron neutrinos (or antineutrinos) and warrants further consideration.  The effects of mixing between sterile neutrinos and the other active neutrino flavors were also not considered.  Further work is required to determine the impact of mixing between a possible sterile neutrino and all of the active neutrino flavors, as well as the dependence of this effect on the supernova progenitor mass and EOS.

\begin{acknowledgments}

The authors wish to thank H. Sam Dalhed (LLNL) for useful discussions regarding the subtleties of running the supernova code.  Work at the University of Notre Dame supported by the U.S. Department of Energy under Nuclear Theory Grant No.~DE-FG02-95-ER40934. 
This work was also supported in part by Grants-in-Aid for Scientific Research of the Japan Society for the Promotion of Science (Grants No.~20105004 and 24340060).

\end{acknowledgments}

\bibliography{sneutrino.bib}

\end{document}